\documentclass[manuscript, gmdd, hvmath]{copernicus}
\usepackage{natbib}
\usepackage{booktabs}           
\usepackage[redefsymbols=false,
            detect-weight = true,
            detect-inline-weight=math]{siunitx}        
\DeclareSIUnit{\year}{a}
\DeclareSIUnit{\month}{mo}
\DeclareSIUnit{\molar}{M}
\DeclareSIUnit{\sverdrup}{Sv}
\DeclareSIUnit{\dpm}{dpm}

\usepackage[dvipsnames]{xcolor}
\definecolor{darkgreen}{RGB}{0,100,0}       
\usepackage{subfig}

\DeclareRobustCommand{\DutchName}[4]{#2~#1}

\begin{document}\hack{\sloppy}
\def\introductionexists{true}
\def\conclusionsexists{true}

\nolinenumbers              

\title{A global scavenging and circulation ocean model of thorium-230 and
protactinium-231 with realistic particle dynamics (NEMO--ProThorP 0.1)
\footnote{In peer review for Geosci.\ Model\ Dev. (discussion paper at \url{https://doi.org/10.5194/gmd-2017-274}).}}
\Author[1,2]{Marco}{van~Hulten}
\Author[1]{Jean-Claude}{Dutay}
\Author[1]{Matthieu}{Roy-Barman}
\affil[1]{Laboratoire des Sciences du Climat et de l'Environnement, IPSL, CEA--Orme des Merisiers, 91191 Gif-sur-Yvette, France}
\affil[2]{Geophysical Institute, University of Bergen, Bergen, Norway}

\runningtitle{Global ocean model of \chem{^{230}Th} and \chem{^{231}Pa}}
\runningauthor{M.M.P.~van~Hulten et~al.}
\correspondence{M.~M.~P.~van~Hulten (\texttt{Marco.Hulten@uib.no})}


\firstpage{1}
\maketitle

\hrulefill

\begin{abstract}
In this paper, we set forth a 3-D ocean model of the radioactive trace isotopes
\chem{^{230}Th} and~\chem{^{231}Pa}.
The interest arises from the fact that these isotopes are extensively used for
investigating particle transport in the ocean and reconstructing past ocean
circulation.
The tracers are reversibly scavenged by biogenic and lithogenic particles.

Our simulations of \chem{^{230}Th} and \chem{^{231}Pa} are based on the NEMO--\textsc{Pisces} ocean
biogeochemistry general circulation model, which includes biogenic particles,
namely small and big particulate organic carbon, calcium carbonate and biogenic
silica.
Small and big lithogenic particles from dust deposition are included in our
model as well.
Their distributions generally compare well with the small and big lithogenic
particle concentrations from recent observations from the \textsc{Geotraces} programme, except
for boundary nepheloid layers for which, as up to today, there are no
non-trivial, prognostic models available on a global scale.
Our simulations reproduce \chem{^{230}Th} and \chem{^{231}Pa} dissolved concentrations: they compare
well with recent \textsc{Geotraces} observations in many parts of the ocean.
Particulate \chem{^{230}Th} and \chem{^{231}Pa} concentrations are significantly improved compared to
previous studies, but they are still too low because of missing particles from
nepheloid layers.
Our simulation reproduces the main characteristics of the \chem{^{231}Pa}/\chem{^{230}Th} ratio observed
in the sediments, and supports a moderate affinity of \chem{^{231}Pa} to biogenic silica as
suggested by recent observations, relative to \chem{^{230}Th}.

Future model development may further improve understanding, especially when
this will include a more complete representation of all particles, including
different size classes, manganese hydroxides and nepheloid layers.
This can be done based on our model, as its source code is readily available.
\end{abstract}

\introduction                           \label{sec:part_gmd:intro}

Oceanic circulation and carbon cycle play a major role in the regulation of the
past and present climate.
Heat and carbon dioxide in the atmosphere tend to equilibrate with the ocean
surface, and are transported down into the deep ocean through the Meridional
Overturning Circulation (MOC\@).
Biogeochemical cycling also generates organic carbon that transfers into the
deep ocean through particle sinking.
Because of this, the strength of the MOC and particle removal participate
actively in the regulation of the climate on the Earth.

Trace elements are also affected by these mechanisms, and represent useful tools
to provide constraints on these processes.
The \textsc{Geotraces} programme has generated a unique, large data set that can
now be used to better understand biogeochemical oceanic processes.
Modelling quantifies and provides more information on the processes that control
the oceanic distribution of these new observations.
In present day climate, it is difficult to measure the MOC strength, and for
past climate there are no measurements available at all.
Isotopes and trace elements from sediment cores are used as proxies to infer past ocean
circulation.
Several examples include carbon isotopes \citep{broecker1988}, the
cadmium/calcium ratio \citep{rosenthal1997}, the ratio between protactinium-231
and thorium-230 (\chem{^{231}Pa}/\chem{^{230}Th}) \citep{boehm2015}, and the neodymium isotope ratio
\chem{^{143}Nd}/\chem{^{144}Nd} \citep{piotrowski2005}.
These proxies are affected by dynamical and biogeochemical processes.
Including these proxies in a climate model is a way to better understand the
climatic signal they register.

We will focus on \chem{^{230}Th} and \chem{^{231}Pa}, because these isotopes are well documented by the
international \textsc{Geotraces} programme, and they are particularly suitable to
study the transfer of particulate matter since the isotopes' source in the ocean is
perfectly known: radioactive decay of uranium isotopes.
Others have modelled \chem{^{143}Nd} and
\chem{^{144}Nd} \citep{arsouze2008,arsouze2009,ayache2016}, \chem{^{13}C}
\citep{tagliabue2009:carbon} and \chem{^{14}C} \citep{mouchet2013}.
The ratio \chem{^{230}Th} and \chem{^{231}Pa} is used as a proxy for past ocean conditions, but this
signal is potentially affected by both circulation and biogeochemical changes.
Therefore, a correct understanding of the scavenging and the underlying particle
dynamics is essential in order to better simulate these tracers
\citep{dutay2015}.

Protactinium-231 and thorium-230 are produced in the ocean by the $\alpha$-decay
of uranium-235 and uranium-234, respectively.
Because the activity of uranium is approximately uniform in the ocean,
\chem{^{231}Pa} and
\chem{^{230}Th} are produced at a relatively constant rate ($\beta_\chem{Pa} =
\SI{2.33e-3}{\dpm\per\cubic\metre\per\year}$ and $\beta_\chem{Th} =
\SI{2.52e-2}{\dpm\per\cubic\metre\per\year}$, dpm = \emph{disintegrations
per minute}) \citep{henderson2003}.
They are both scavenged rapidly by the many particles that reside in the ocean
and settle towards the seafloor.
\chem{^{231}Pa} is less sensitive to particle scavenging than \chem{^{230}Th}, which is reflected in the
longer residence time of \chem{^{231}Pa} (80--200\;yr) compared to that of \chem{^{230}Th} (20--40\;yr)
\citep{yu1996}.
\chem{^{231}Pa} and \chem{^{230}Th} are radioactive, decaying to radium isotopes and having a half-life
of 32.76\;kyr and 75.40\;kyr, respectively.
Each combination of particle--radionuclide adsorption has a different
reactivity.
The vertical distributions of natural radionuclides, such as \chem{^{230}Th} and
\chem{^{231}Pa}, are
hence sensitive to the distribution and mixture of particles.
As a consequence of the different particle reactivities of \chem{^{230}Th} and
\chem{^{231}Pa},
$\chem{[^{231}Pa]}_D/\chem{[^{230}Th]}_D$ deviates from the production activity
ratio of 0.093 \citep{anderson1983:open,anderson2003,rutgersvanderloeff2016}.

As both particle dynamics and circulation of the ocean affect \chem{^{230}Th}
and \chem{^{231}Pa},
numerical biogeochemical general circulation models are used to study the relative contribution of these mechanisms.
The isotopes have been simulated in models of intermediate complexity,
for instance by \citet{henderson1999} (LSG-OGCM), \citet{marchal2000} (EMIC\;2.5D), \citet{siddall2007}
(EMIC\;3D), \citet{heinze2006} (HAMOCC\@) and \citet{luo2010}.
 More complex ocean general circulation models models have also been used to simulate these tracers
 \citet{dutay2009} (NEMO--\textsc{Pisces}) and
\citet{rempfer2017} (Bern3D\@).
\citet{dutay2009} demonstrated that the particle concentration simulated by the
\textsc{Pisces} model in the deep ocean was much too low.
This lead to overestimated radionuclide concentrations in the deep ocean.
Therefore, it is crucial to improve the representation of the particles
\citep{dutay2009}.
\citet{rempfer2017} showed that taking into account additional sinks at the
seafloor and at the ocean margins yields an improved agreement with
observations, especially for the dissolved phases of \chem{^{230}Th} and
\chem{^{231}Pa}.
Particulate ratios improved to a lesser extent; the authors have not
presented evaluation of their simulated particulate concentrations.

In this study, we try to improve on previous studies that simulate the
distribution of \chem{^{230}Th} and \chem{^{231}Pa}.
Our approach is to improve the mechanistic description of the particle and
radionuclide cycling, as well as on the side of system design and reusability of
the model.
We evaluate how well the model fits with observations, but tuning is not
one of our main goals.
Just like \citet{dutay2009}, we use the NEMO--OPA ocean general circulation model and
the \textsc{Pisces} biogeochemical model with improved particle dynamics \citep{aumont2017}.
There are several improvements since \citet{dutay2009}, namely the new model
\begin{itemize}
\item includes lithogenic particles from dust deposition;
\item improved the biogeochemistry, affecting the biogenic particle distributions \citep{aumont2015,aumont2017};
\item includes three different phases per nuclide (dissolved, and adsorbed
onto big (and small) particles), whereas \citet{dutay2009} include only a
single compartment (total concentration) for each nuclide from which they
calculated the respective phases based on chemical equilibrium;
\item is more precise/explicit on the mathematical formalism;
\item is written in Fortran 95 instead of Fortran 77, making the extension of
its use to other modern Fortran models easier;
\item is part of a modern model framework, NEMO--TOP, facilitating its use with
other models in this framework.
\end{itemize}
\noindent
\citet{aumont2017} showed that dissolution rates of Particulate Organic Carbon
(POC) were overestimated, and they improved on this by introducing a spectrum of
different labilities.
This improved the simulation of both small and large POC significantly, so we
use this same model for our simulations.
Furthermore, we develop a model of lithogenic particles
that also scavenge \chem{^{230}Th} and~\chem{^{231}Pa}.
The main objective of this paper is to improve on the simulation of
\chem{^{230}Th} and \chem{^{231}Pa}
(the dissolved and two particulate size classes of particles for both nuclides),
based on a more realistic modelling of small and big particles.

New observations are available from the \textsc{Geotraces} programme.
Especially the North Atlantic GA03 transects will be used for model validation;
because on this transect not only dissolved \chem{^{230}Th} and \chem{^{231}Pa} have been measured
\citep{hayes2015:scavenging}, but also their adsorbed forms
\citep{hayes2015:scavenging} and biogenic and lithogenic particle
concentrations in two size classes \citep{lam2015:size}.

\section{Model description}             \label{sec:part_gdm:methods}

In order to simulate the biogenic particle dynamics and its interaction with the
\chem{^{230}Th} and \chem{^{231}Pa} trace isotopes, we use the biogeochemical circulation model
NEMO--\textsc{Pisces} \citep{madec2008,aumont2015}.
This model has been employed for many other studies concerning trace metals, as
well as large-scale ocean biogeochemistry
\citep[e.g.][]{gehlen2007,arsouze2009,dutay2009,tagliabue2010,vanhulten:alu_jms,vanhulten:alu_bg,vanhulten:mang_bg}.
We force \textsc{Pisces} by a climatological year of circulation fields
(including turbulent diffusion) that was obtained from the dynamical
component of the \textit{Nucleus for European Modelling of the Ocean} (NEMO\@).
Table~\ref{tab:model} gives an overview of this and other components of the
model and their relevant properties.

\begin{table*}
\centering
\begin{tabular}{llr}
\toprule
model component     & improvements or relevant properties   & timestep   \\
\midrule
circulation (OPA)   & prescribed forcing (off-line physics) \citep[e.g.\ used by][]{arsouze2009} & 6.0\;h \\
\addlinespace
biogenic particles  & prognostically integrated by \textsc{Pisces} \citep{aumont2015} & 1.5\;h \\
 -- POC             & two size classes settling with $w_s =
\SI{2}{\metre\per\day}$ and $w_b = \SI{50}{\metre\per\day}$; variable reactivity \citep{aumont2017} & \\
 -- \chem{CaCO_3}   & one size class settling with $w_b = \SI{50}{\metre\per\day}$; higher order dissolution \citep[e.g.][]{subhas2015} & \\
 -- \chem{bSiO_2}   & one size class settling with $w_b = \SI{50}{\metre\per\day}$; no changes & \\
\addlinespace
lithogenic particles& based on forcing of dust deposition; two size classes
settling with $w_s = \SI{2}{\metre\per\day}$ and $w_b = \SI{50}{\metre\per\day}$ & 1.5\;h \\
\addlinespace
radionuclide model        & six prognostic tracers (dissolved, small and large
adsorbed; \chem{^{231}Pa} and \chem{^{230}Th}) & 6.0\;h \\
\bottomrule
\end{tabular}
\caption{Model components and properties essential for the radionuclide model.}
\label{tab:model}
\end{table*}

All model fields are defined on the ORCA2 discrete coordinate system, an
irregular grid covering the whole world ocean with a nominal resolution of
$2\degree\!\times 2\degree$, with an increased resolution in the meridional
direction near the equator and Antarctica, and in both horizontal directions in
the Mediterranean, Red, Black and Caspian Seas.
On the Northern Hemisphere, it has two coordinate singularities,
one in Canada and the other in Russia, such that both singularities
fall outside the computational domain.
The vertical resolution of the ORCA2 grid is 10\;m in the upper 100\;m, increasing
downwards to 500\;m, such that there are 30 layers in total and the ocean has a
maximum depth of 5000\;m \citep{madec1996,murray1996}.
The timestep of the model is 6\;h for the dynamics and the radionuclides, and
1.5\;h for the biogeochemistry (PISCES) and the lithogenic particles.
When necessary, sub-timestepping is done for all sinking components in the model.

\subsection{Circulation}

The circulation was obtained by forcing NEMO in the ORCA2 configuration with
air-sea boundary conditions, consisting of heat, fresh water and momentum fluxes
that were derived from bulk formulae.
They are functions of wind, sea surface temperature, air temperature, air
humidity and evaporation minus precipitation.
For the tropics, daily wind stress was used, which was based on European
Remote Sensing (ERS) satellite data, and for the polar regions, NCEP/NCAR
re-analysis data \citep{kalnay1996,kistler2001}.
Surface salinity was restored with a timescale of 60~days towards the seasonal
Polar Science Center Hydrographic Climatology (PHC) dataset to avoid model drift
\citep{timmermann2005}.
The last year of this simulation is used as our one-year climatology with a
resolution of five days of the dynamics.

%

\subsection{Particle dynamics}

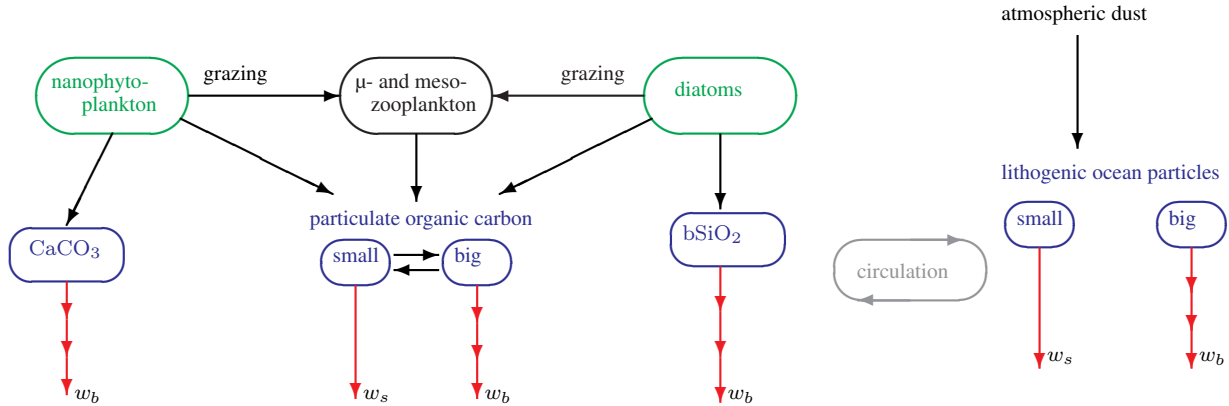
\begin{figure*}
    \setlength{\unitlength}{1.0mm}
    {\footnotesize
    \begin{picture}(160,80)
        \thicklines
        \put(8,70){\color{Green}\oval(20,10)}
        \put(0,71){\color{Green}nanophyto-}
        \put(4,68){\color{Green}plankton}
        \put(20,72){grazing}            
        \put(18,70){\vector(1,0){20}}   
        \put(17,67){\vector(2,-1){20}}  
        \put(8,65){\vector(-1,-2){6}}
        \put(2,49){\color{Blue}\oval(15,7)}
        \put(-3,49){\color{Blue}\chem{CaCO_3}}
        \multiput(2,35.5)(0,5){3}{\color{Red}\vector(0,-1){5}}
        \put(3,30){$w_b$}

        \put(48,70){\color{Black}\oval(20,10)}
        \put(40,71){\color{Black}\unit{\mu}- and meso-}
        \put(42,68){\color{Black}zooplankton}
        \put(48,65){\vector(0,-1){9}} 
        \put(40,48){\color{Blue}\oval(9,6)}
        \put(37,48){\color{Blue}small}
        \put(45,49){\vector(1,0){6}}
        \put(51,47){\vector(-1,0){6}}
        \put(56,48){\color{Blue}\oval(9,6)}
        \put(53,48){\color{Blue}big}
        \put(34,53){\color{Blue}particulate organic carbon}
        \multiput(56,35)(0,5){3}{\color{Red}\vector(0,-1){5}}
        \put(41,30){$w_s$}
        \put(40,45){\color{Red}\vector(0,-1){15}}
        \put(57,30){$w_b$}

        \put(88,70){\color{Green}\oval(20,10)}
        \put(82,70){\color{Green}diatoms}
        \put(67,72){\color{Black}grazing}
        \put(78,70){\color{Black}\vector(-1,0){20}}
        \put(79,67){\vector(-2,-1){20}} 
        \put(88,65){\vector(0,-1){10}}
        \put(89,51){\color{Blue}\oval(15,7)}
        \put(83,51){\color{Blue}\chem{bSiO_2}}
        \multiput(88,35.5)(0,6){3}{\color{Red}\vector(0,-1){6}}
        \put(89,30){$w_b$}

        \put(125,80){atmospheric dust}
        \put(135,78){\vector(0,-1){15}}
        \put(125,59){\color{Blue}lithogenic ocean particles}
        \put(130,53){\color{Blue}\oval(9,6)}
        \put(127,53){\color{Blue}small}
        \put(150,53){\color{Blue}\oval(9,6)}
        \put(147,53){\color{Blue}big}
        \put(130,50){\color{Red}\vector(0,-1){16}}
        \put(131,35){$w_s$}
        \multiput(150,40)(0,5){3}{\color{Red}\vector(0,-1){6}}
        \put(151,35){$w_b$}

        \put(113,47){\color{Gray}\oval(20,8)}
        \put(113,51){\color{Gray}\vector(1,0){7}}
        \put(113,43){\color{Gray}\vector(-1,0){7}}
        \put(106,46){\color{Gray}circulation}
    \end{picture}
    }
    \caption{Conceptual model of particle dynamics, on which the numerical model
    is based.
    Nanophytoplankton and diatoms (in {\color{Green}green}) take up nutrients and \chem{CO_2},
    which are released again from respiration and remineralisation of
    \chem{bSiO_2}, POC, DOM and lithogenic particles.
    The nutrients are not represented in the figure, because only particles
    impact \chem{^{231}Pa} and \chem{^{230}Th}.
    Zooplankton are denoted by the {\color{black}black} box.
    All sinking particles are denoted by {\color{Blue}blue} boxes.
    Effectively, this figure comprises the internal cycling of \textsc{Pisces},
    minus details that are not of interest here, plus the lithogenic dust model.
    DOM stands for Dissolved Organic Carbon, sPOM and bPOM stand for small and big
    Particulate Organic Matter which are both subject to the differential
    lability scheme \citep{aumont2017}, and \chem{bSiO_2} stands for biogenic silica.
    Sinking is denoted by the {\color{Red}red} arrows; tripple arrows means fast, normal arrows slow.
    }
    \label{fig:particle_dynamics}
\end{figure*}

This version of \textsc{Pisces} includes two size classes of POC, both with
differential remineralisation rates \citep{aumont2017}, one class of biogenic
silica (\chem{bSiO_2}) and one class of calcium carbonate (\chem{CaCO_3}).
In the model, particles sink down with two velocities:
$w_s = \SI{2}{\metre\per\day}$ and $w_b = \SI{50}{\metre\per\day}$, corresponding with ``small''
and ``big'' particles (Fig.~\ref{fig:particle_dynamics}).

For \chem{CaCO_3} we changed the standard first-order dissolution kinetics
parameterisation to a fourth-order dissolution.
For our simulation we used a calcite dissolution rate constant of
$k=\SI{2.5}{\per\month}$ and a
dissolution order of $n=3.9$ \citep{subhas2015}.

\begin{figure}
\centering
\includegraphics[width=\linewidth]{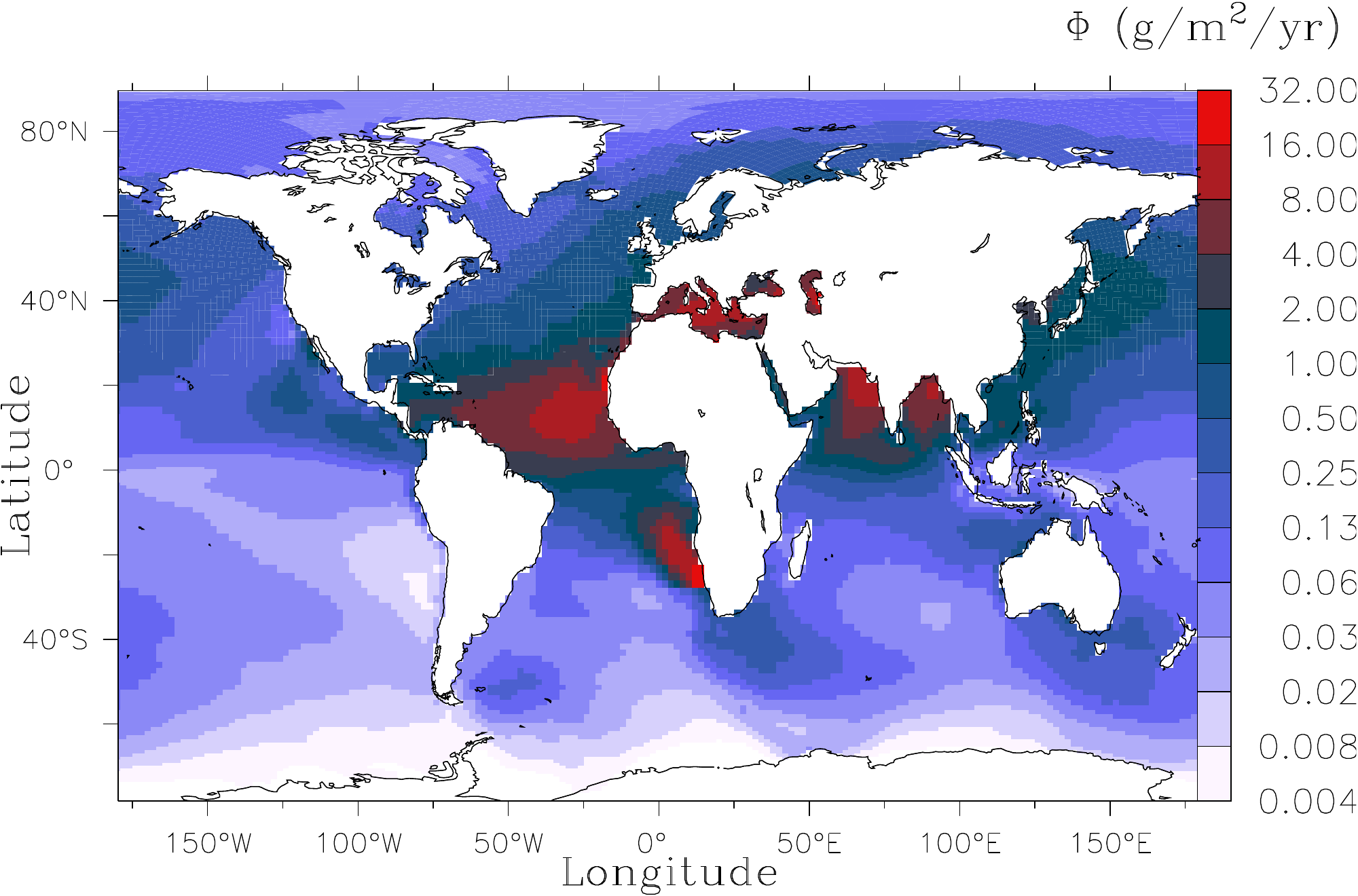}
\caption{Dust deposition on a logarithmic scale (\si{\gram\per\square\metre\per\year}).
It is the integrated flux over the 12-months climatology based on \citet{hauglustaine2004}.}
\label{fig:dust}
\end{figure}

In addition to biogenic particles, we introduced lithogenic dust particles
in the model.
The yearly average dust flux is derived from \citet{hauglustaine2004} and is
presented in Fig.~\ref{fig:dust}.
It is used by the model as the input of lithogenic ocean particles as well as
for nutrient supply in \textsc{Pisces}.
This dust deposition field has been tested in biogeochemical studies with this
configuration of NEMO \citep[e.g.]{vanhulten:alu_jms}.

Small and large lithogenic particles are added to the upper layer of the ocean
according to:
\begin{equation}
    \frac{\partial P^\mathrm{Litho}}{\partial t}\Big|_\mathrm{surface}
      = \frac{f}{\Delta z_1} \cdot \Phi_\mathrm{dust} \,,
\end{equation}
where $P^\mathrm{Litho}$ is the small (large) lithogenic particle concentration,
$f$ is the fraction of the dust that gets partitioned into the small (large)
lithogenic particles in the ocean, $\Delta z_1 = \SI{10}{\metre}$ is the
thickness of the upper model layer, and $\Phi_\mathrm{dust}$ is the dust flux.

Our model has two size classes for lithogenic particles, so this equation is
applied for two different concentrations $P^\mathrm{Litho}$ and respective fractions~$f$.
We set the small lithogenic dust flux fraction to 20\,\% and the big one to~80\,\%.
Once partitioned in the ocean, the lithogenic particles sink down, changing
their concentrations throughout the ocean according to
\begin{equation}
    \frac{\mathrm{d} P^\mathrm{Litho}}{\mathrm{d} t}
        = - w \cdot \frac{\partial P^\mathrm{Litho}}{\partial z}
          + \big(\mathcal{A}\nabla_{\!h}^2 + \mathcal{B}\frac{\partial^2}{\partial z^2}\big) P^\mathrm{Litho} \,,
    \label{eqn:settling}
\end{equation}
where $w$ is the settling velocity, set to the constant \SI{2}{\metre\per\day}
for the small lithogenic particles and to \SI{50}{\metre\per\day} for the big
particles.
The depth, $z$, is positive upwards, $\nabla_{\!h}$ is the horizontal
divergence, and $\mathcal{A}$ and~$\mathcal{B}$
are respectively the horizontal and vertical eddy diffusivity coefficients.
The material derivative includes a term for eddy-induced velocity
\citep{gent1990,gent1995}.
Equation~\eqref{eqn:settling} is identical to the settling of small and big POC
\citep{aumont2015}.
Of course, there are also biological and chemical sources and sinks for POC\@.
However, for lithogenic particles there are no such sources or sinks because the
only source in our model is dust deposition and we assume the lithogenic
particles are refractory.
The lithogenic particles are removed from the model domain when arriving at the
seafloor, which means that they are buried in the sediment.

\subsection{Radionuclides}


Thorium-230 and protactinium-231 are produced throughout the ocean from the
decay of \chem{^{234}U} and \chem{^{235}U}, respectively.
Because of long residence times of over 200\;kyr, these uranium isotopes are
approximately homogeneously distributed throughout the ocean, and do not change much over
time \citep{ku1977}.
The residence time of \chem{^{234}U} is 3.2--\SI{5.6e5}{\year}, which is much
longer than the full mixing time of the world ocean of about \SI{e3}{\year} \citep{dunk2002}.
The \chem{^{234}U} concentration can vary about 10\,\%, depending mostly on the salinity
\citep{owens2011}, but that is smaller than uncertainties arising from other
assumptions in our model.
Therefore, we assume that the production rates of \chem{^{230}Th}
and \chem{^{231}Pa} are constant, both in space and time.

\begin{figure}
    \centering
    \setlength{\unitlength}{0.6mm}
    {\footnotesize
    \begin{picture}(120,60)
        \thicklines
        \put(-5,40){\vector(1,0){36}}
        \put(-5,35){in situ production}

        \put(50,46){\color{Gray}\vector(2,1){18}}
        \put(108,46){\color{Gray}\vector(-2,1){18}}
        \put(65,58){\color{Gray}decay products}

        \put(45,20){\color{Gray}\oval(30,12)}
        \put(45,26){\color{Gray}\vector(1,0){8}}
        \put(45,14){\color{Gray}\vector(-1,0){8}}
        \put(35,19){\color{Gray}circulation}

        \put(48,40){\color{Orange}\oval(29,13)}
        \put(36,41){\color{Orange}dissolved}
        \put(36,37){\color{Orange}radioisotope}
        \put(68,45){adsorption}
        \put(66,42){\vector(1,0){26}}
        \put(92,37){\vector(-1,0){26}}
        \put(68,31){desorption}
        \put(108,40){\color{Blue}\oval(26,12)}
        \put(100,39){\color{Blue}particle}
        \put(108,34){\color{Red}\vector(0,-1){25}}
        \put(111,21){settling}
    \end{picture}
    }
    \caption{The conceptual reversible scavenging model for the radionuclides.
    The radioisotopes, \chem{^{230}Th} and \chem{^{231}Pa}, are depicted in {\color{Orange}orange} when in the dissolved phase.}
    \label{fig:scavenging}
\end{figure}
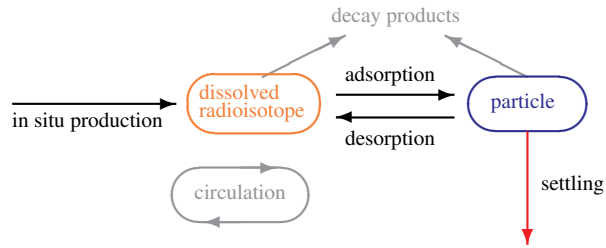

The \chem{^{230}Th} and \chem{^{231}Pa} radionuclides are reversibly scavenged by biogenic and
lithogenic particles (Fig.~\ref{fig:scavenging}).
We will assume, as in previous studies \citep[e.g.][]{siddall2005,dutay2009}, that the adsorption and desorption
reaction rates are much faster than radionuclide production, decay, advection,
mixing, change in particle distribution and settling of the adsorbed phases.
Because of that, we equilibrate between the dissolved and adsorbed phases
instantly at each time step.
This means, only considering ad- and desorption processes at this point,
that we must solve this set of equations for every nuclide~$i$:
\begin{subequations}\label{eqn:sorption}
\begin{align}
    0 &= \frac{\partial}{\partial t}(A_{i,D} + A_{i,S} + A_{i,B})\,,\label{eqn:a}\\
    A_{i,S} &= A_{i,D} \sum_{j\in S} K_{ij}P^j   \,,        \label{eqn:b}\\
    A_{i,B} &= A_{i,D} \sum_{j\in B} K_{ij}P^j   \,,        \label{eqn:c}
\end{align}
\end{subequations}
where $A_i$ stands for the activity of nuclide $i \in \{\chem{^{230}Th},
\chem{^{231}Pa}\}$,\footnote{%
Concentration of radionuclides (amount per unit of volume) is proportional to
its (radio)activity (disintegrations per unit of time).
These terms are used interchangeably throughout this paper.
Moreover, they are considered identical: the activity or concentration
$\chem{[^{230}Th]}_D \equiv A_\chem{^{230}Th}$ is expressed in
\si{\milli\becquerel\per\cubic\metre}, and thus we report activity ratios of
\chem{^{231}Pa}/\chem{^{230}Th}.}
and where $P^j$ stands for the concentration of particle $j \in \{\mathrm{sPOC},
\mathrm{bPOC}, \chem{CaCO_3}, \chem{bSiO_2}, \mathrm{sLitho}, \mathrm{bLitho}\}$
(in gram per gram of seawater, so strictly this is a \emph{mass fraction}).
$D$ is the set of non-sinking phases (here only dissolved), $S$ of small
particles that sink with \SI{2}{\metre\per\day} and $B$ of big particles that
sink with \SI{50}{\metre\per\day} (in the same way as Eq.~\eqref{eqn:settling}
for lithogenic particles).
For any particle size class $J\in \{S,B\}$ and any radionuclide $i$, we define
$A_{i,J} = \sum_{j\in J} A_{ij}$.
Finally, $K_{ij}$ is the equilibrium partition coefficient of nuclide $i$ for
particle~$j$.

Since we cannot solve this analytically, this will be done numerically by first
assigning a new value to the dissolved nuclide activity:\footnote{%
When summation bounds are not specified, the union of all phases, $D \cup S
\cup B$, is assumed.}
\begin{equation}
    A_{i,D} := \frac{A_{i,D} + A_{i,S} + A_{i,B}}{1 + \sum_j K_{ij}P^j} \,.
\end{equation}
Then, we calculate the activity of $i$ that is adsorbed onto small and big
particles by applying Eqs~\eqref{eqn:b} and~\eqref{eqn:c}.
With this approach, the small and big adsorbed concentrations equilibrate
instantly.
Assuming that the change in adsorption strength is much smaller than
the relative change in tracer activity, $\forall_{i, J:} \partial_t \sum_{j\in J}
K_{ij} P^j \ll \partial_t A_{i,D} / A_{i,D}$,
the total activity of every $i$ is conserved, i.e.\ Eq.~\eqref{eqn:a} holds.
\begin{proof}
  Let the total adsorption strength for any isotope $i$ be $Q_i = \sum_j K_{ij}P^j$
  and the total amount of the same isotope $T_i = A_{i,D}+A_{i,S}+A_{i,B}$, and let primes
  denote the updated concentrations.
  Assume that the adsorption strength for every isotope $i$ is constant ($Q_i' \equiv Q_i$).
  \begin{align*}
    T_i' &= \frac{T_i}{1+Q_i'} + A_{i,D}' Q_i' \\
    &= \frac{T_i}{1+Q_i} + \frac{T_i}{1+Q_i} Q_i \\
    &= \frac{T_i}{1+Q_i} (1+Q_i) = T_i \\
    \Rightarrow \partial_t T_i' &= \partial_t T_i = 0 \,. &&\qedhere
  \end{align*}
\end{proof}
\noindent

The \chem{^{230}Th} and \chem{^{231}Pa} that are adsorbed onto the particles follow the same law as the
small and big (lithogenic) particles (Eq.~\ref{eqn:settling}).
Of course, by definition, the adsorbed radioisotope and the particle settle with
the same speed, and thus we have implemented it.

The decay terms of \chem{^{230}Th} and \chem{^{231}Pa} are much smaller than the other sources
and sinks, but they are included in the model:
\begin{subequations}
\begin{align}
\frac{\mathrm{d}A_{i,D}}{\mathrm{d}t} &= \beta_i-\lambda_iA_{i,D}
+ (\mathcal{A}\nabla_{\!h}^2 + \mathcal{B}\frac{\partial^2}{\partial z^2}) A_{i,D}   \,,\\
\frac{\mathrm{d}A_{i,S}}{\mathrm{d}t} &= -\lambda_i A_{i,S}-w_s \frac{\partial A_{i,S}}{\partial z}
+ (\mathcal{A}\nabla_{\!h}^2 + \mathcal{B}\frac{\partial^2}{\partial z^2}) A_{i,S}    \,,\\
\frac{\mathrm{d}A_{i,B}}{\mathrm{d}t} &= -\lambda_i A_{i,B}-w_b \frac{\partial A_{i,B}}{\partial z}
+ (\mathcal{A}\nabla_{\!h}^2 + \mathcal{B}\frac{\partial^2}{\partial z^2}) A_{i,B}    \,,
\end{align}
\end{subequations}
where $\beta_i$ and $\lambda_i$ are respectively the production rate and
radioactive decay of isotope~$i$.

\section{Simulations}

The model was spun up for 400\;yr, after which it was in an
approximate steady state (decadal drift of $-0.002$\,\% for total
\chem{^{230}Th} and $+0.056$\,\% for total \chem{^{231}Pa}).
Protactinium-231 has a larger drift than thorium-230, because \chem{^{230}Th}
is everywhere in the ocean more quickly removed because of its high particle
reactivity.
The lithogenic particles are in a steady state, and the PISCES variables are in
an approximate steady state (e.g.\ phosphate shows a drift of $-0.005$\,\% per
decade.


The partition coefficients $K_{ij}$, with $i \in \{\chem{Th}, \chem{Pa}\}$,
depend on the type of particle $j$ and are given in Table~\ref{tab:K_d}.
\begin{table}
    \centering
    \begin{tabular}{lr@{.}lr@{.}ll}
        \toprule
        Particle        & \multicolumn{2}{l}{$K_\chem{Pa}$} (\si{\mega\gram\per\gram})
                                        & \multicolumn{2}{l}{$K_\chem{Th}$}
                                            (\si{\mega\gram\per\gram})
                                                        & Settling speed  \\
        \midrule
        small POC       &   2&0         &   5&0         & \SI{2}{\metre\per\day} \\
        big POC         &   0&4         &   1&0         & \SI{50}{\metre\per\day} \\
        biogenic silica &   0&5 (0.4)   &   0&5 (1.0)   & \SI{50}{\metre\per\day} \\
        \chem{CaCO_3}   &   0&12        &   5&0         & \SI{50}{\metre\per\day} \\
        small lithogen. &  10&0         &  50&0         & \SI{2}{\metre\per\day} \\
        large lithogen. &   1&0         &   5&0         & \SI{50}{\metre\per\day} \\
        \bottomrule
    \end{tabular}
    \caption{Partition coefficients for the different modelled particles.
    Between braces is the value for the sensitivity simulation discussed in
    Section~\ref{sec:opal}.}
    \label{tab:K_d}
\end{table}
These coefficients have large uncertainties still, but their values can be
constrained by reported values from different experimental studies
\citep[e.g.][]{chase2002,geibert2004,hayes2015:scavenging}.
Therefore, we had quite some freedom in prescribing values.
The adsorption onto calcium carbonate is a factor two decreased from
\citet{chase2002}.
This brings the $K$ value of \chem{Pa} closer to that in
\citet{hayes2015:scavenging} based on field measurements ($0.9\pm
0.4$\;\si{\mega\gram\per\gram}), but since we maintained the ratio of $K$
values, the partition coefficient of \chem{Th} is much smaller than found by
\citet{hayes2015:scavenging}.
However, it is consistent with the laboratory results of \citet{geibert2004}.
For small lithogenic particles, $K$ is set about a factor five larger than literature
\citep{geibert2004,hayes2015:scavenging}, whereas large lithogenic particles have a smaller
value than reported in the literature.
The consequences of chosen values of the partition coefficients is discussed
further in Section~\ref{sec:discussion}.

\section{Observations}

In this study, we will focus on the \textsc{Geotraces} GA03 transect in the North
Atlantic Ocean.
Recently, a large amount of measurements on both radionuclides and their carrier
particles have been collected at this transect.
This unique combination makes it especially useful to evaluate a radionuclide
scavenging model.
We will also compare our global ocean model with several observational datasets
throughout the ocean (Table~\ref{tab:data}).
Observations obtained from the \textsc{Geotraces} programme are denoted with the
respective \textsc{Geotraces} transect number \citep{mawji2015}.

\begin{table*}[t]
  \begin{tabular}{lllll}
    \toprule
    Transect    & Year      & Expedition                    & Ocean basin           & Citation                  \\
    \midrule
    \multicolumn{4}{l}{Carrier particles POC, \chem{CaCO_3}, \chem{bSiO_2} and lithogenic}  \\
    \midrule
    --          & 1973--74  & GEOSECS st.235,239,306        & Central Pacific           & \citet{bruncottan1991}    \\
    --          & 1985--91  & Alcyone-5, Eve-1, Hydros-6    & North Pacific             & \citet{druffel1992}       \\
    --          & 1982--97  & WCR, Line P, SOFeX, K2, JGOFS & Pacific and Atlantic      & \citet{lam2011}           \\
    GA03        & 2011      & US GT10 and GT11              & North Atlantic            & \citet{lam2015:size}      \\
    \addlinespace
    \midrule
    \multicolumn{4}{l}{Radionuclides \chem{^{230}Th} and \chem{^{231}Pa}}  \\
    \midrule
    --          & 1979      & R/V~\textit{Knorr} cruise 73/leg 16    & Central-east Pacific      & \citet{bacon1982}         \\
    --          & 1994      & R/V~\textit{Moana Wave}, HOT-57        & Central Pacific           & \citet{roybarman1996}     \\
    --          & 1976--98  & (multiple)                    & global                    & \citet{henderson1999}         \\
    --          & 1983--02  & (multiple)                    & global                    & \citet{marchal2007}       \\
    GIPY5\_w    & 2008      & ANT XXIV/3                    & Southern Ocean / Drake    & \citet{venchiarutti2011}  \\
    GIPY5\_e    & 2008      & ANT XXIV/3                    & Southern Ocean / Zero merid.& \citet{venchiarutti2011}  \\
    --          & 2009      & SO202-INOPEX, ALOHA, SAFe     & North Pacific             & \citet{hayes2013:lithogenic}\\
    GA02        & 2011      & JC\,057                       & Southwest Atlantic Ocean  & \citet{deng2014}          \\
    GA03        & 2010      & US GT10 and GT11              & North Atlantic            & \citet{hayes2015:scavenging}\\
    \addlinespace
    \midrule
    \multicolumn{4}{l}{Sediment top core \chem{^{231}Pa}/\chem{^{230}Th}}  \\
    \midrule
    --          & 1966--97  & (multiple)                    & global                    & \url{http://climotope.earth.ox.ac.uk/data_compilations/holocene_231pa230th_dataset_notes_and_references}\\
    --          & 2009      & R/V~\textit{M.~Dufresne}, MD173/RETRO3 & global           & \citet{burckel2015,burckel2016}\\
    \bottomrule
  \end{tabular}
  \caption{Observations used for comparison with the model simulations.}
  \label{tab:data}
\end{table*}

For the carrier particles most of our data come from \citet{lam2015:size},
which is a recent \textsc{Geotraces} dataset at the GA03 transect in the North Atlantic Ocean.
An older compilation of particles is taken from \citet{lam2011}.
We use some older data as well (Table~\ref{tab:data}).

Concentrations of both dissolved and particulate phases of \chem{^{230}Th} and
\chem{^{231}Pa} were
taken from \citet{hayes2015:thpa} (GA03) and \citet{hayes2014:biogeography}
(Pacific Ocean).
Other data are listed in Table~\ref{tab:data}.

To evaluate the sediment \chem{^{231}Pa}/\chem{^{230}Th} flux of the model, we will compare with
compilations of the Holocene (i.e.\ top core particulate concentrations)
(Table~\ref{tab:data}).

\section{Results}


\subsection{Circulation}

As mentioned before, our model is part of a global general circulation model,
but instead of solving the Navier-Stokes equations, our tracers are advected by
the circulation fields of a previous simulation of the dynamical ocean model.
Since the overturning circulation is of importance for the redistribution of the
tracers, we here present basic results thereof.

\begin{figure} 
    \centering
    \includegraphics[width=\linewidth]{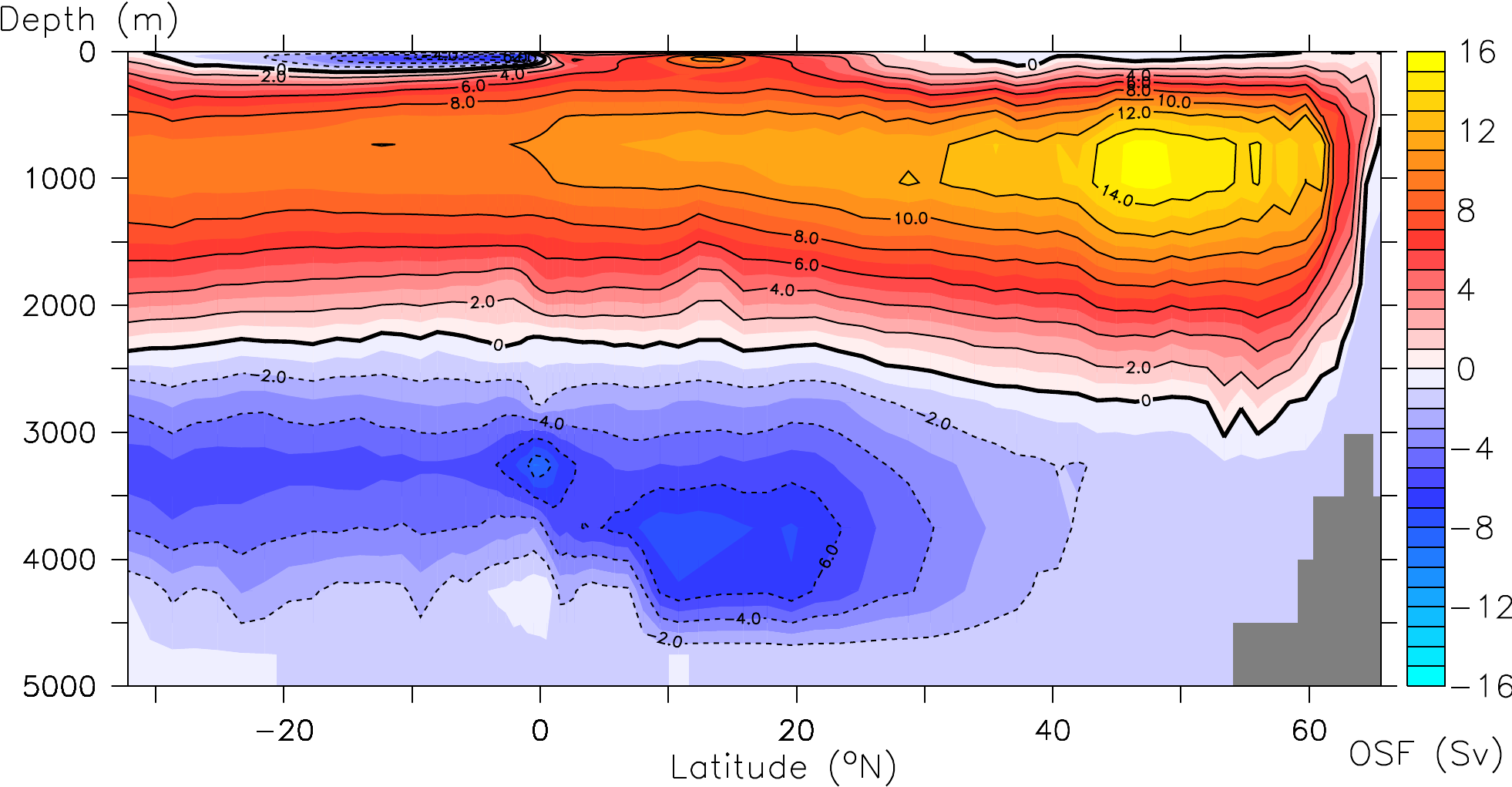}
    \caption{The Atlantic Overturning Stream Function (OSF), a measure for the
             AMOC in Sv (1\;Sv = $10^6$\;\si{\cubic\metre\per\second}; clockwise
             is positive).}
    \label{fig:osf}
\end{figure}

Figure~\ref{fig:osf} presents the Overturning Stream Function (OSF) of the
Atlantic Ocean.
The OSF is defined as the zonally (through the basin) and vertically (from the
surface downwards) integrated meridional current speed.
We use this as a measure for the Atlantic Meridional Overturning Circulation
(AMOC\@).
The upper overturning cell transports \SI{14}{\sverdrup}
($\SI{1}{\sverdrup} = \SI{e6}{\cubic\metre\per\second}$) on average.
Observations suggest higher values of about $19 \pm 5$\;Sv
\citep{talley2003,cunningham2007,rayner2011}.
The lower cell has a strong overturning strength of about \SI{6}{\sverdrup}, which is
mostly due to the fact that the AMOC is shallow.
The AMOC reaches about \SI{2500}{\metre} depth, whereas observational studies
show the deep water sink down to about \SI{4000}{\metre}.
At around 30\degree\,N the AMOC has weakened notably, which is consistent with
some studies \citep[e.g.][]{johnson2008}, but the AntArctic Bottom Water (AABW)
does not go as far north as other studies suggest
\citep[e.g.~45\degree\;N in][]{vanaken2011}.

The Antarctic Circumpolar Current is also an important feature for large-scale
ocean circulation.
The through-flow at Drake Passage is a good measure for that.
In our model the flux through Drake Passage is 200\;Sv, which slightly
overestimates recently reported values \citep[e.g.\ $173\pm
11$\;Sv in][]{donohue2016}.

\subsection{Particles}

\begin{figure*}
\centering
\includegraphics[width=\linewidth]{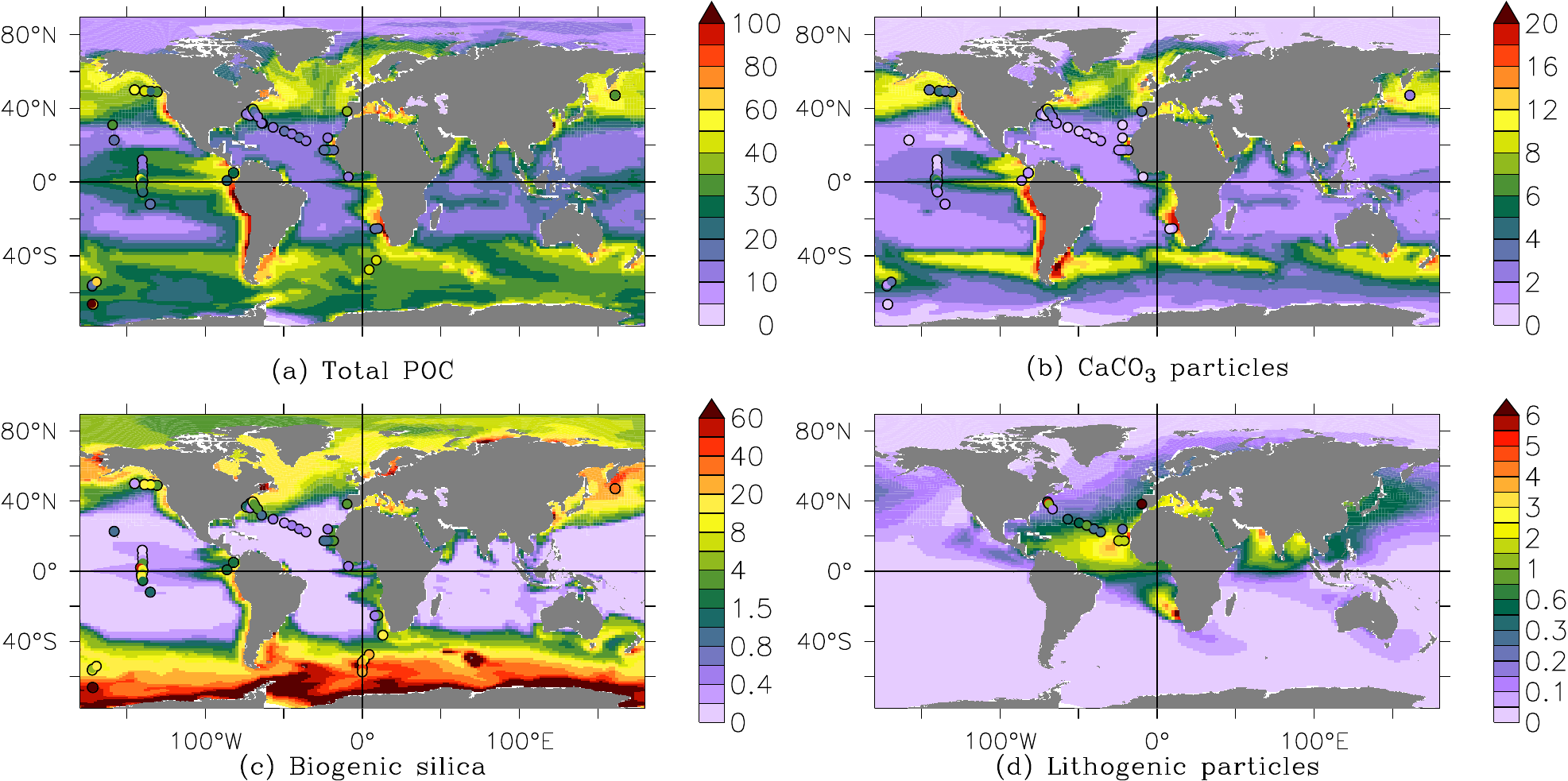}
\caption{Particle concentrations from the \textsc{Pisces} model at the surface
ocean.  Observations are represented as coloured dots.  Units of all modelled
and measured particle concentrations are in \si{\milli\gram\per\cubic\metre}.}
\label{fig:part_surface}
\end{figure*}

Figure~\ref{fig:part_surface} shows the surface concentrations of total POC,
large calcium carbonate, large biogenic silica and total lithogenic particles.
Observations are plotted as dots on top of the modelled concentrations.
The modelled biogenic particles include living matter.
For the model, we assume POC includes all phytoplankton and microzooplankton (not
mesozooplankton since it may swim away), calcium carbonate contains the assumed
fraction of \chem{CaCO_3} in the phytoplankton, and opal includes living
diatoms.
The modelled concentrations and patterns of total POC compare well with the
observations (Fig.~\ref{fig:part_surface}a).
For instance, coastal and equatorial Pacific POC concentrations are both in the
model and the observations elevated compared to other regions, albeit that the
spatial extent of the oligotrophic regions appears to be too small in the model.
We present large \chem{CaCO_3} particles, because we have data of that for both
the Atlantic and Pacific Ocean, whereas from the Atlantic Ocean we only have
small particle data.
Furthermore, our model only includes large (fast-sinking) calcium carbonate
particles.
As the model tries to represent \chem{CaCO_3} with only one size class, neither
comparing with only big particles nor comparing with total \chem{CaCO_3} is
completely fair.
Whereas the meridional patterns of large \chem{CaCO_3} particles are reproduced,
its concentrations are generally overestimated, especially in the Gulf of Alaska
(Fig.~\ref{fig:part_surface}b).
Contrarily, \emph{total} observed \chem{[CaCO_3]}, which includes smaller
particles, of which measurements are available only at GA03, is higher than the
prediction of our modelled (only large) \chem{CaCO_3} particle concentration.
The model produces a realistic spatial distribution of biogenic silica: there are
more elevated values in the high latitudes, but concentrations are
underestimated at the surface at low latitudes (Fig.~\ref{fig:part_surface}c).


\begin{figure*}
    \centering
    \subfloat[Total POC (\si{\micro\molar})]{
        \includegraphics[width=.5\linewidth]{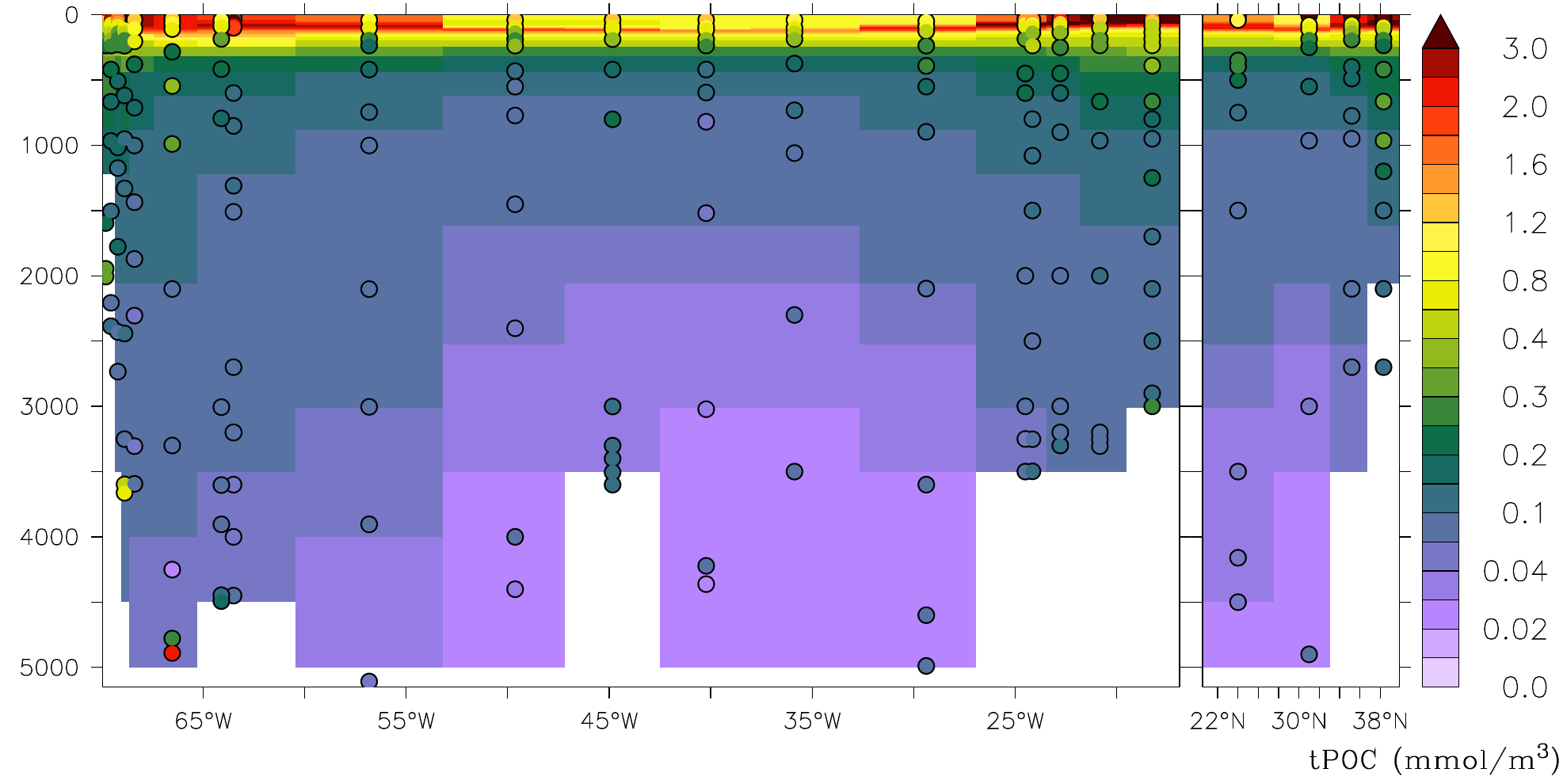}
        \label{fig:tPOC_GA03}
    }
    \subfloat[Fraction of large POC]{
        \includegraphics[width=.5\linewidth]{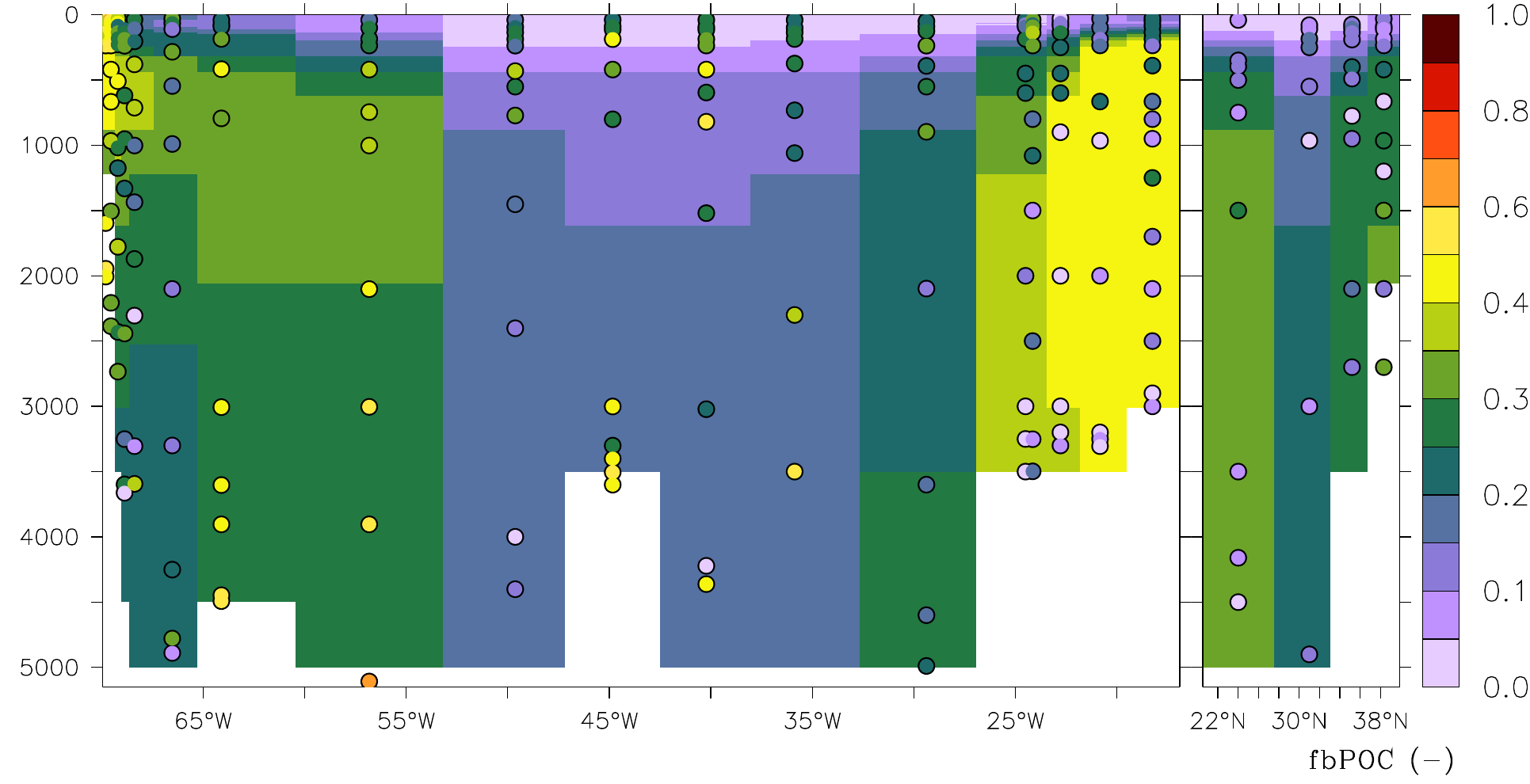}
        \label{fig:fbPOC_GA03}
    }\\
    \subfloat[\chem{CaCO_3} (\si{\micro\molar})]{
        \includegraphics[width=.5\linewidth]{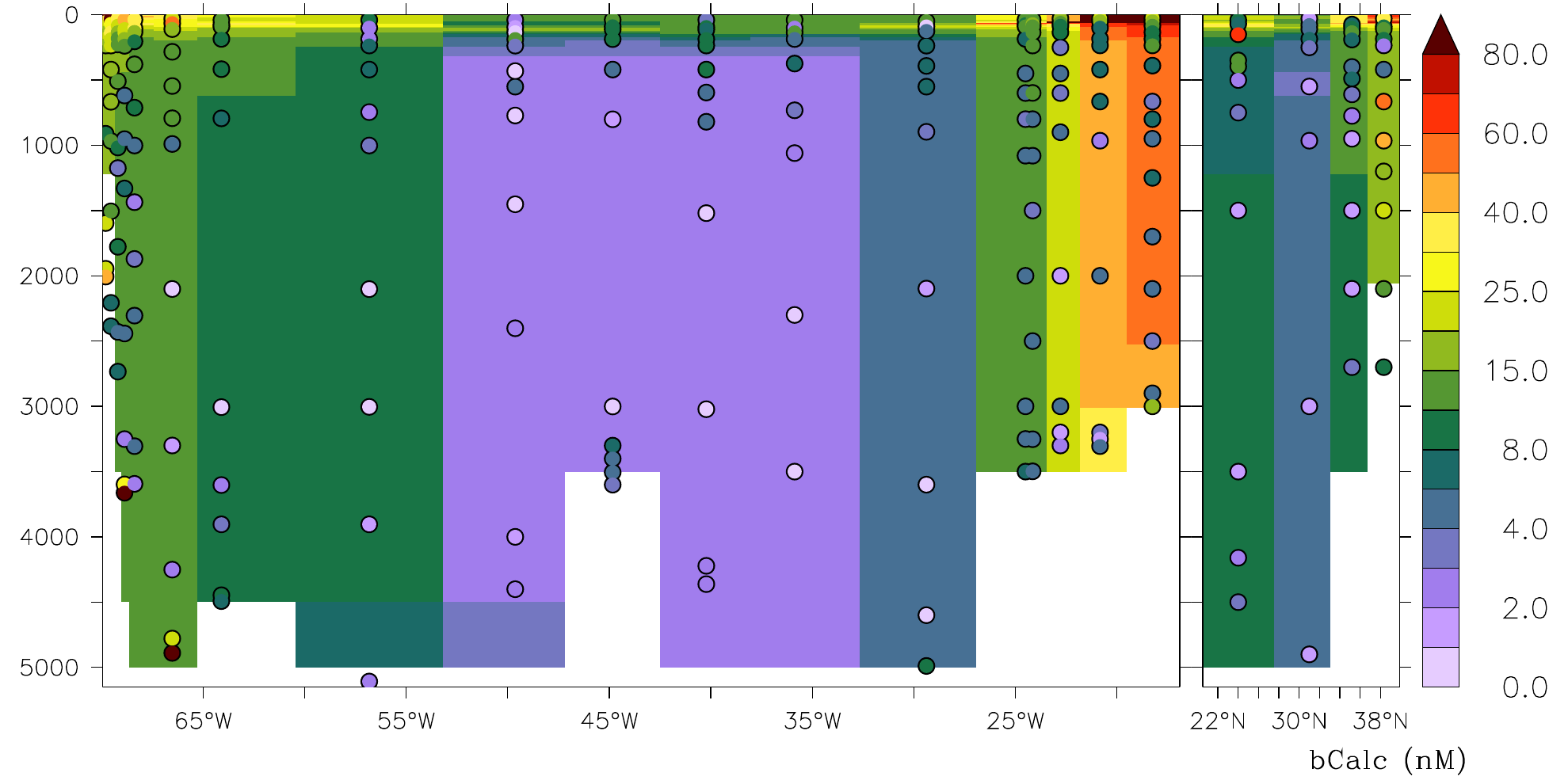}
        \label{fig:bCalc_GA03}
    }
    \subfloat[\chem{bSiO_2} (\si{\micro\molar})]{
        \includegraphics[width=.5\linewidth]{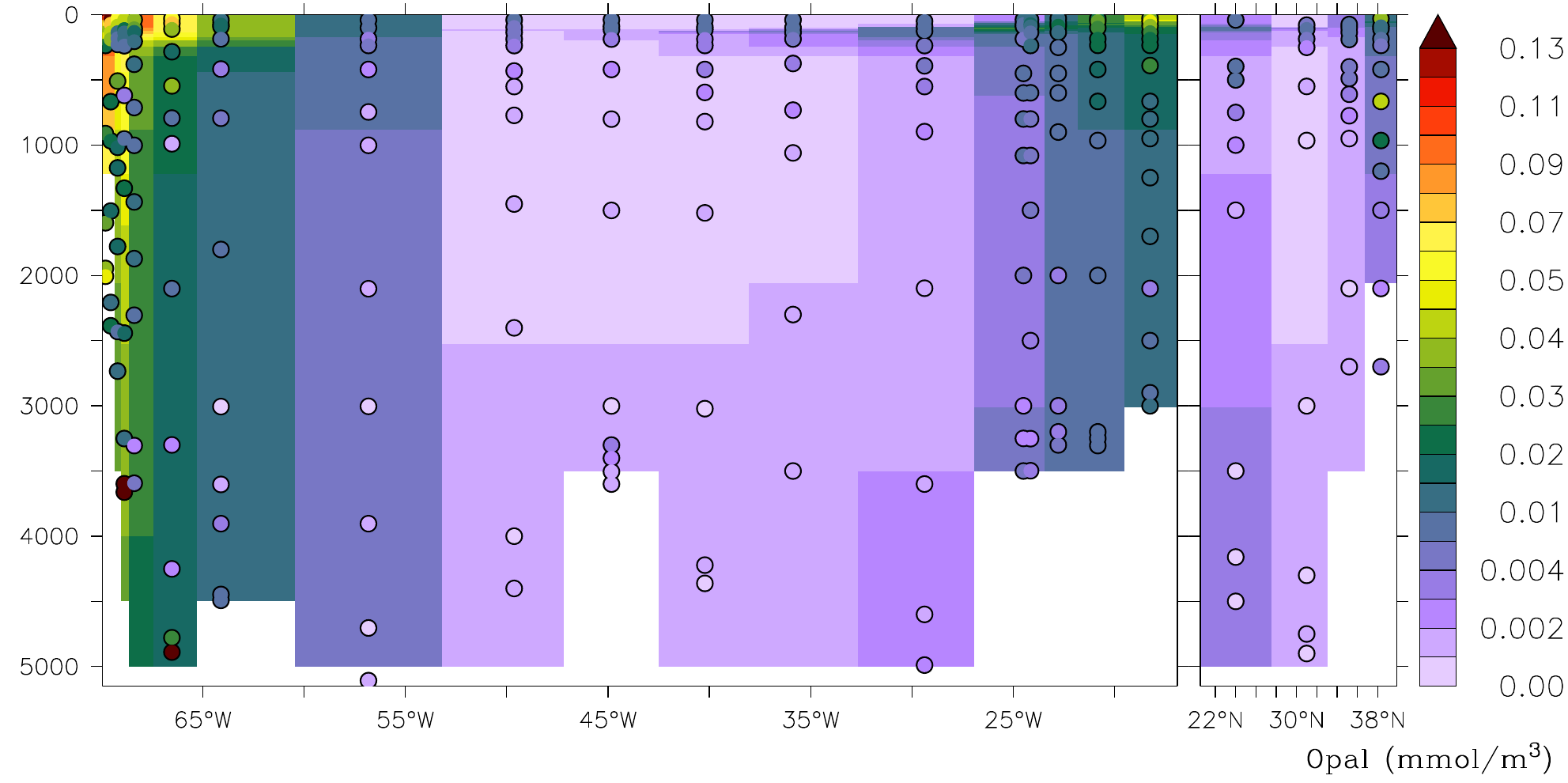}
        \label{fig:Opal_GA03}
    }\\
    \label{fig:bio_GA03}
    \subfloat[Total lithogenic (\si{\milli\gram\per\cubic\metre})]{
        \includegraphics[width=.5\linewidth]{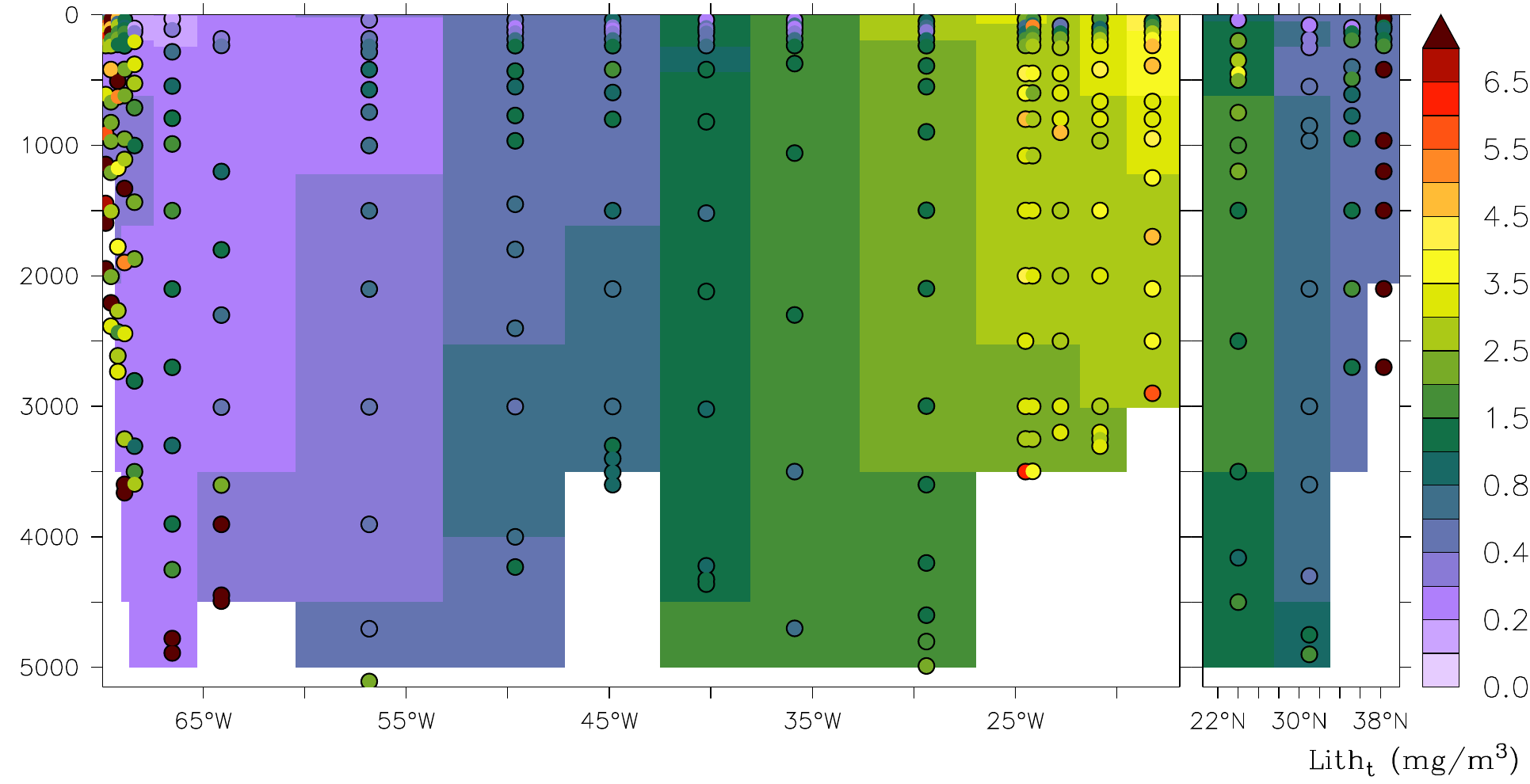}
        \label{fig:Lith_t_GA03}
    }
    \subfloat[Fraction of large lithogenic]{
        \includegraphics[width=.5\linewidth]{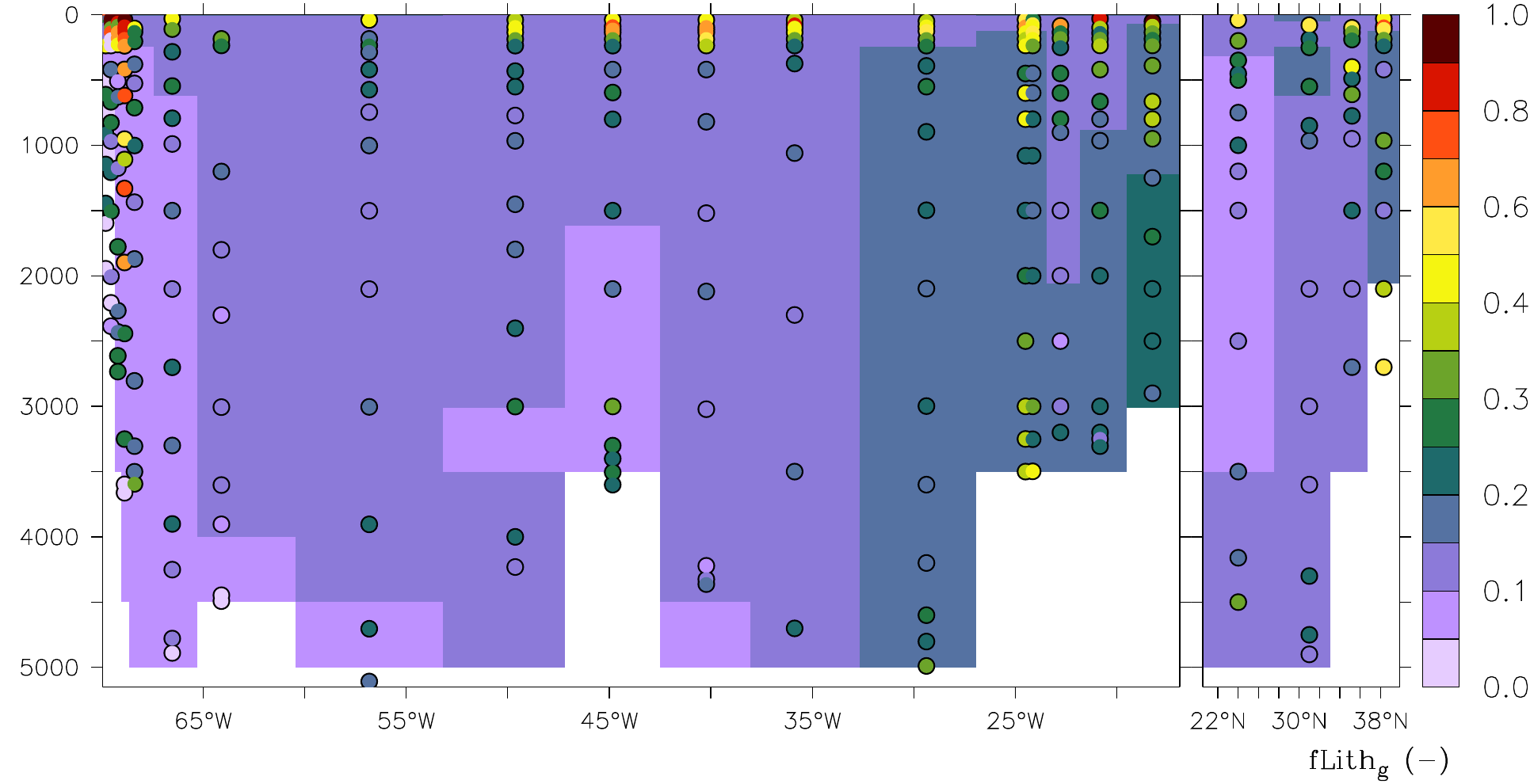}
        \label{fig:fLith_g_GA03}
    }
    \caption{Particle fields at the GA03 North Atlantic \textsc{Geotraces} transect;
            observations \citep{lam2015:size} as coloured dots.}
    \label{fig:part_GA03}
\end{figure*}

The recent large dataset obtained from the full-ocean depth \textsc{Geotraces}
GA03 transect \citep{lam2015:size} provides an opportunity to analyse the
performance of \textsc{Pisces} in more detail.
\textsc{Pisces} generally produces the right order of magnitude for all four
biogenic particles (small and big POC, calcium carbonate and biogenic silica),
but there still remain some shortcomings in their distributions as well
(Fig.~\ref{fig:part_GA03}a--d).
\begin{itemize}
\item The total POC concentration in Fig.~\ref{fig:tPOC_GA03} is up to a
factor of 4 underestimated in the deep oligotrophic ocean.
In the upper \SI{200}{\metre} of the ocean the model overestimates the
observations, though at some points at the surface the POC concentration is
underestimated (also Fig.~\ref{fig:part_surface}a).
In both the observations and the model the fraction of large POC varies from
zero to 0.6 (Fig.~\ref{fig:fbPOC_GA03}), but the spatial distributions are
very different.
More detailed results and discussion on the simulation of POC are to be found in
\citet{aumont2017}.
\item The model underestimates the \chem{CaCO_3} concentrations from
70\degree\,W to 25\degree\,W by a factor of 2 to 10, but east of that up to
Africa the prediction lies close to or overestimates the observations.
In the Canary Basin and up to Portugal (meridional transect at the right),
the model reproduces the right order of magnitude, but it does not reproduce
the correct profiles everywhere (Fig.~\ref{fig:bCalc_GA03}).

\item Biogenic silica concentrations are generally reproduced but the model
overestimates the higher concentrations observed along the western margin
(Fig.~\ref{fig:Opal_GA03}).
\end{itemize}


Finally, the modelled total lithogenic particle concentration at the surface
shows, as expected, a close resemblance with dust deposition patterns
(Fig.~\ref{fig:dust}), and mostly compares well with observations
(Fig.~\ref{fig:part_surface}d).
Only near the coasts of the USA and of Portugal the model strongly
underestimates lithogenic particle concentrations.
Concerning the deep ocean, our model captures quite nicely the general
distribution at the GA03 transect (Figs.~\ref{fig:Lith_t_GA03}
and~\ref{fig:fLith_g_GA03}).
However, the concentrations near the western boundary are strongly
underestimated, especially those of large lithogenic particles.
Our model reproduces the observed fraction of big lithogenic particles of
$\sim$0.1 to about $\sim$0.3 in most of the deep ocean, but it underestimates
the much larger fraction of big particles observed in the upper 200\;m of the
ocean (0.1--0.2 in the model versus 0.3--0.9 in the observations).

\subsection{Thorium-230 and protactinium-231}

\begin{figure*}
\centering
\includegraphics[width=\linewidth]{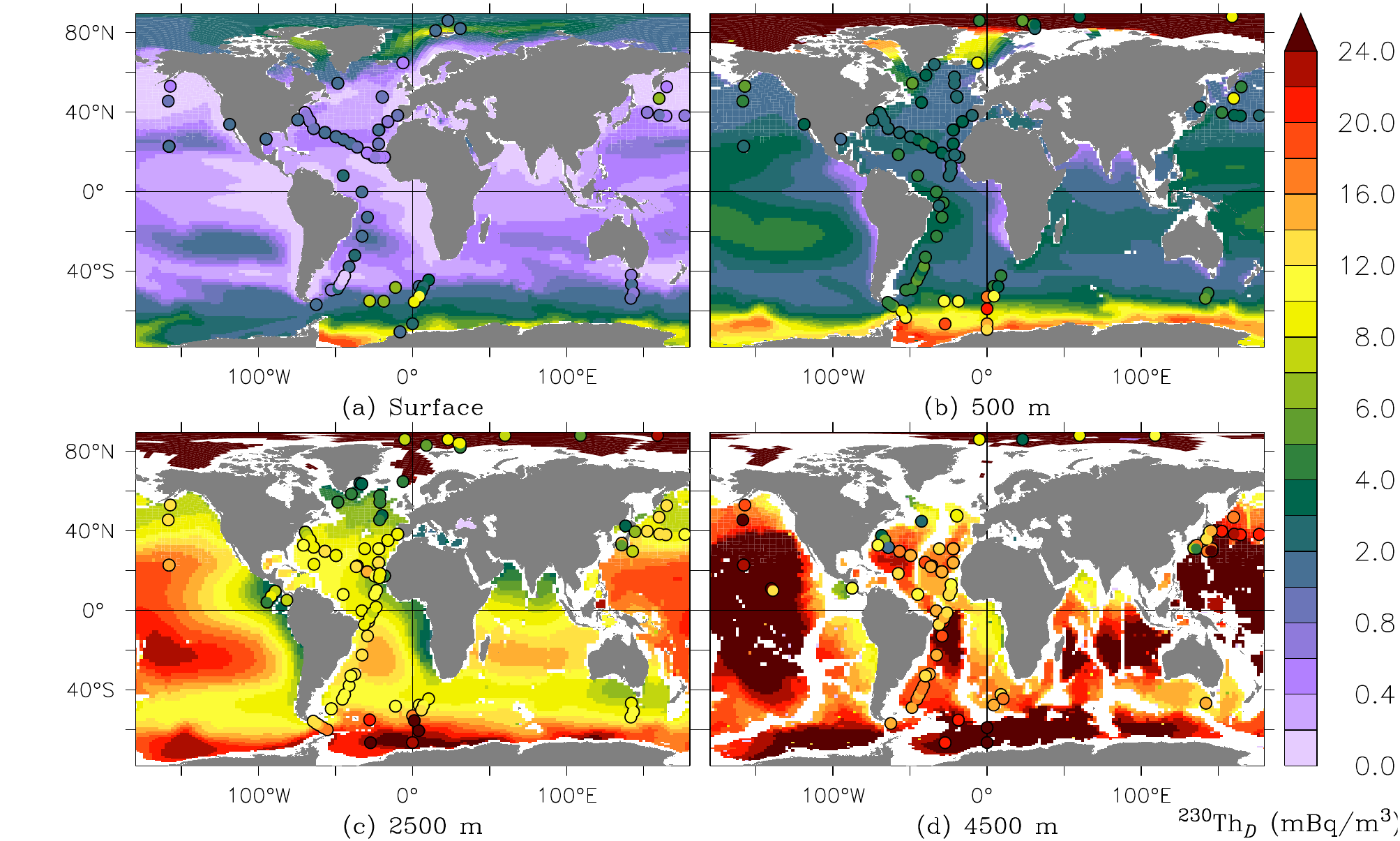}
\caption{The dissolved thorium-230 activity at four depth levels (\si{\milli\becquerel\per\cubic\metre}), observations as dots.}
\label{fig:Th230d_4depths}
\end{figure*}

\begin{figure*}
\centering
\includegraphics[width=\linewidth]{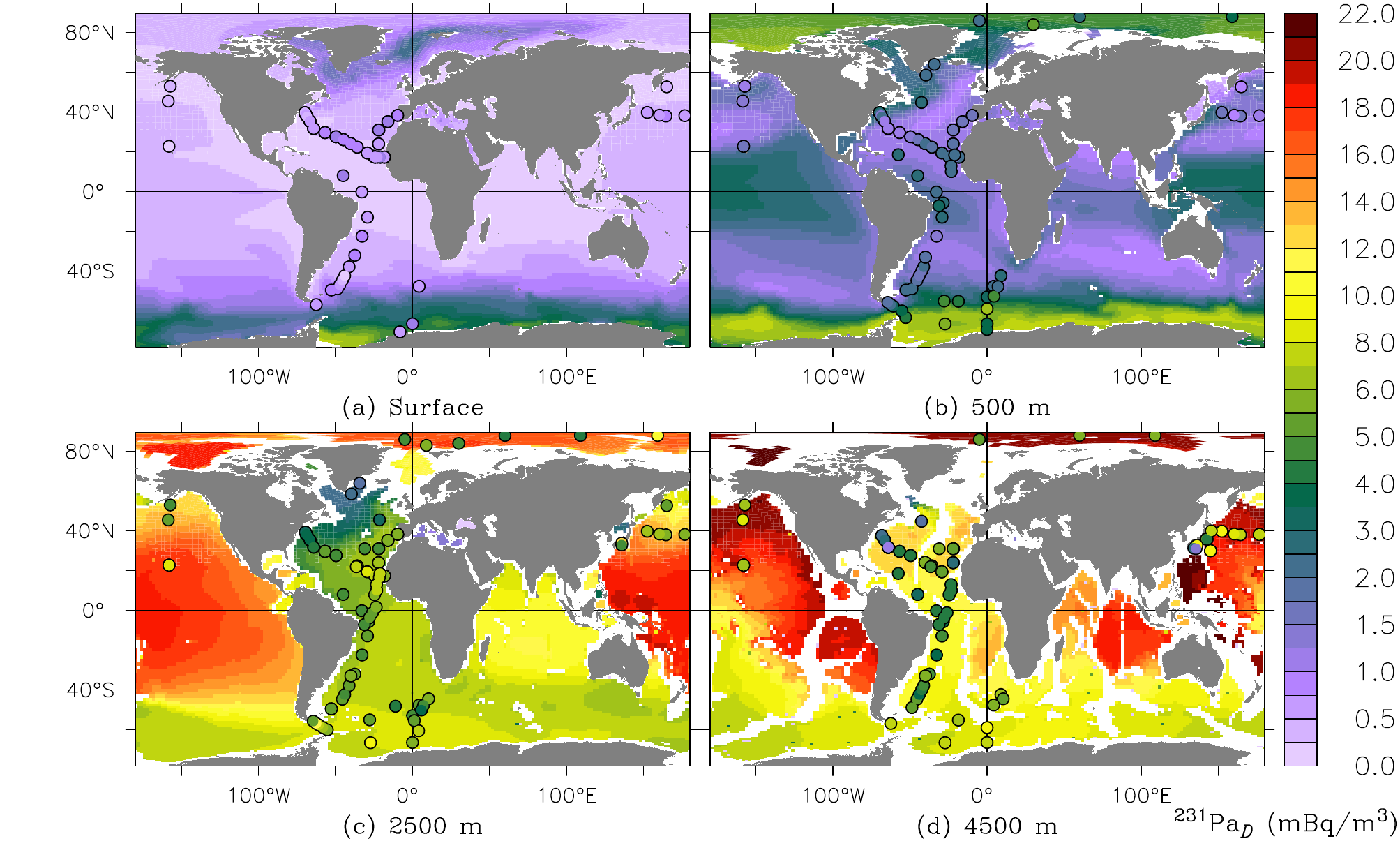}
\caption{The dissolved protactinium-231 activity at four depth levels (\si{\milli\becquerel\per\cubic\metre}), observations as dots.}
\label{fig:Pa231d_4depths}
\end{figure*}

\begin{figure*}
    \centering
    \subfloat[\chem{[^{230}Th]}$_D$]{
        \includegraphics[width=.5\linewidth]{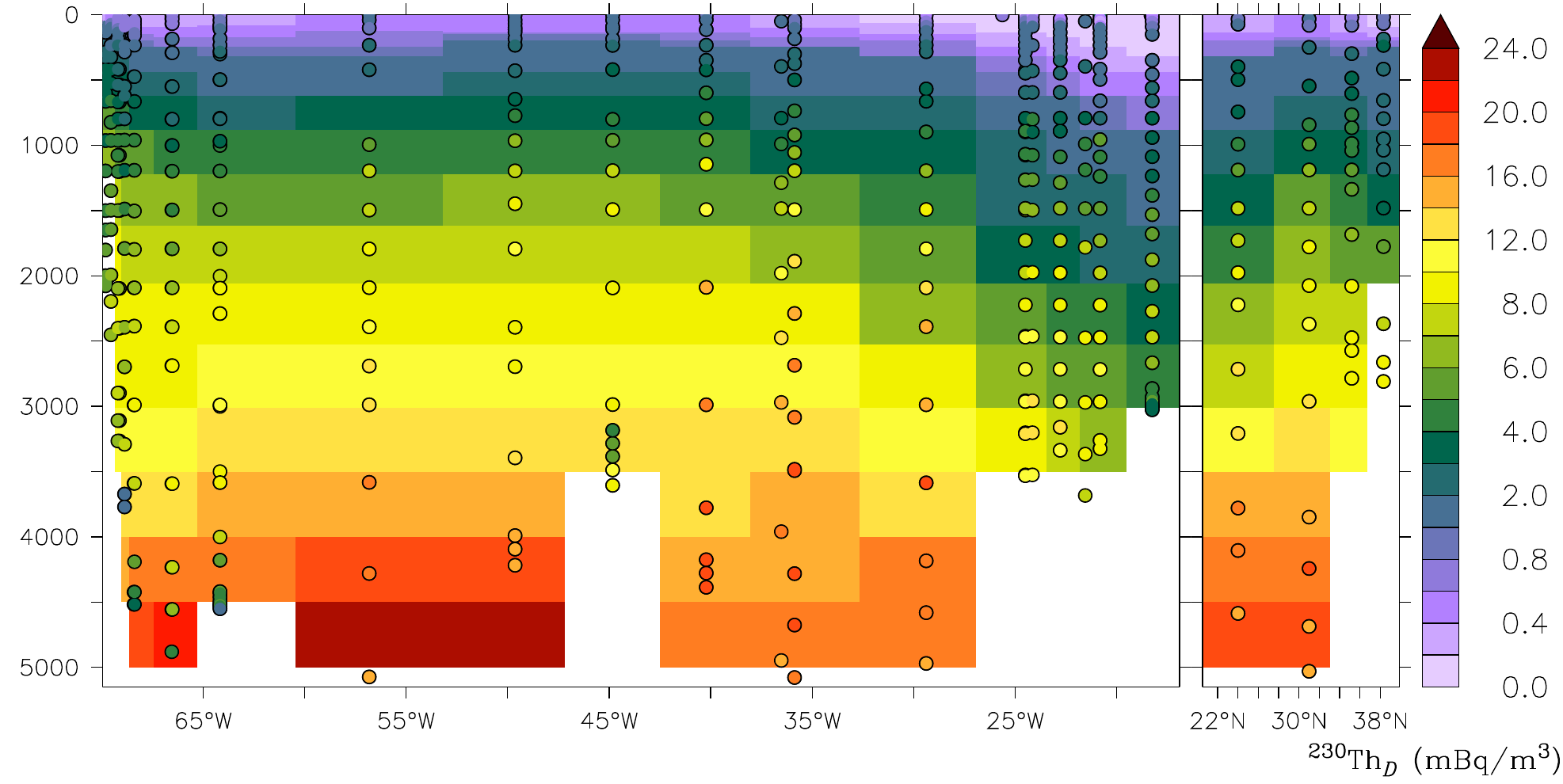}
        \label{fig:Th230d_GA03}
    }
    \subfloat[\chem{[^{231}Pa]}$_D$]{
        \includegraphics[width=.5\linewidth]{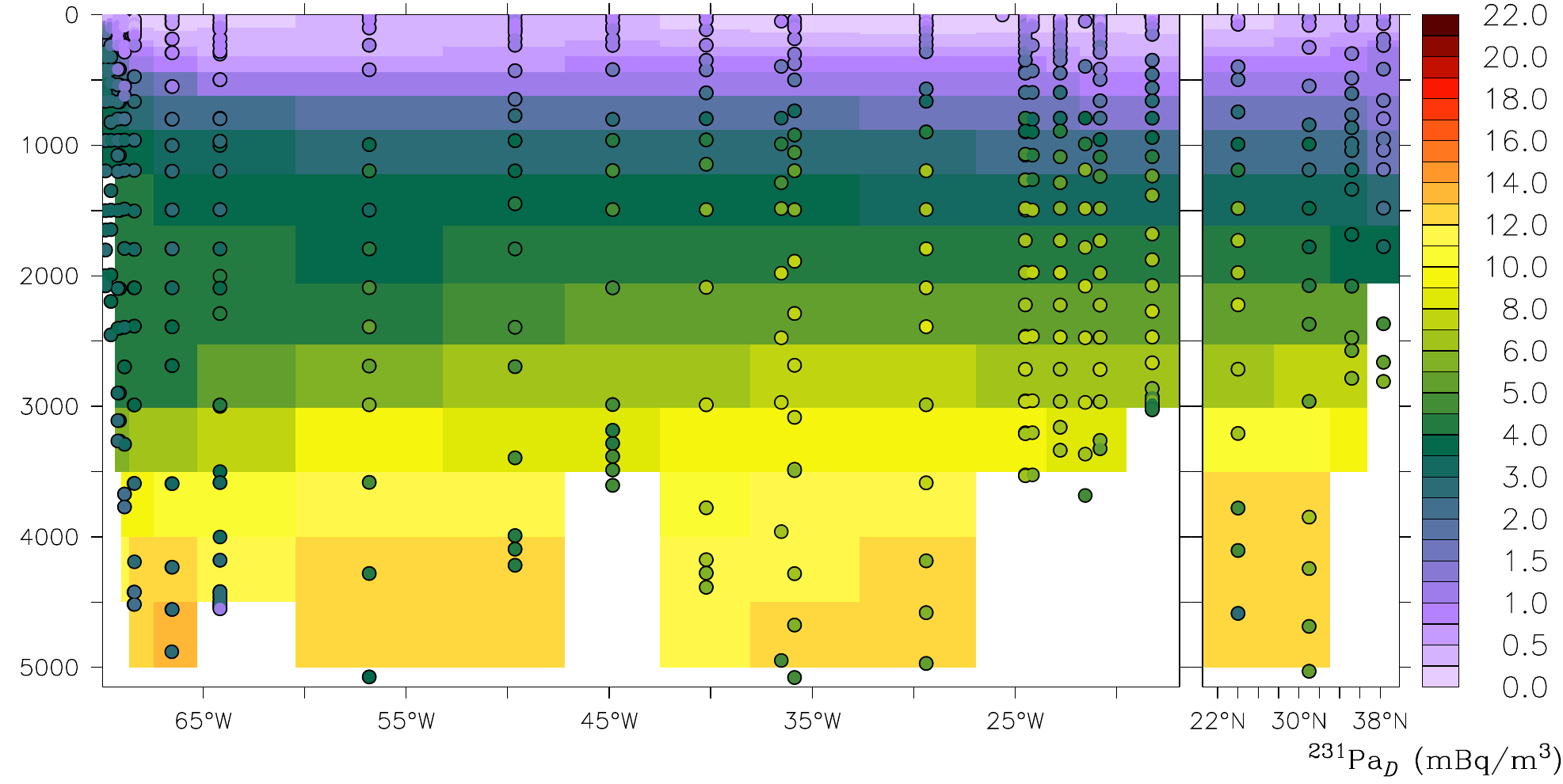}
        \label{fig:Pa231d_GA03}
    }\\
    \subfloat[\chem{[^{230}Th]}$_S$]{
        \includegraphics[width=.5\linewidth]{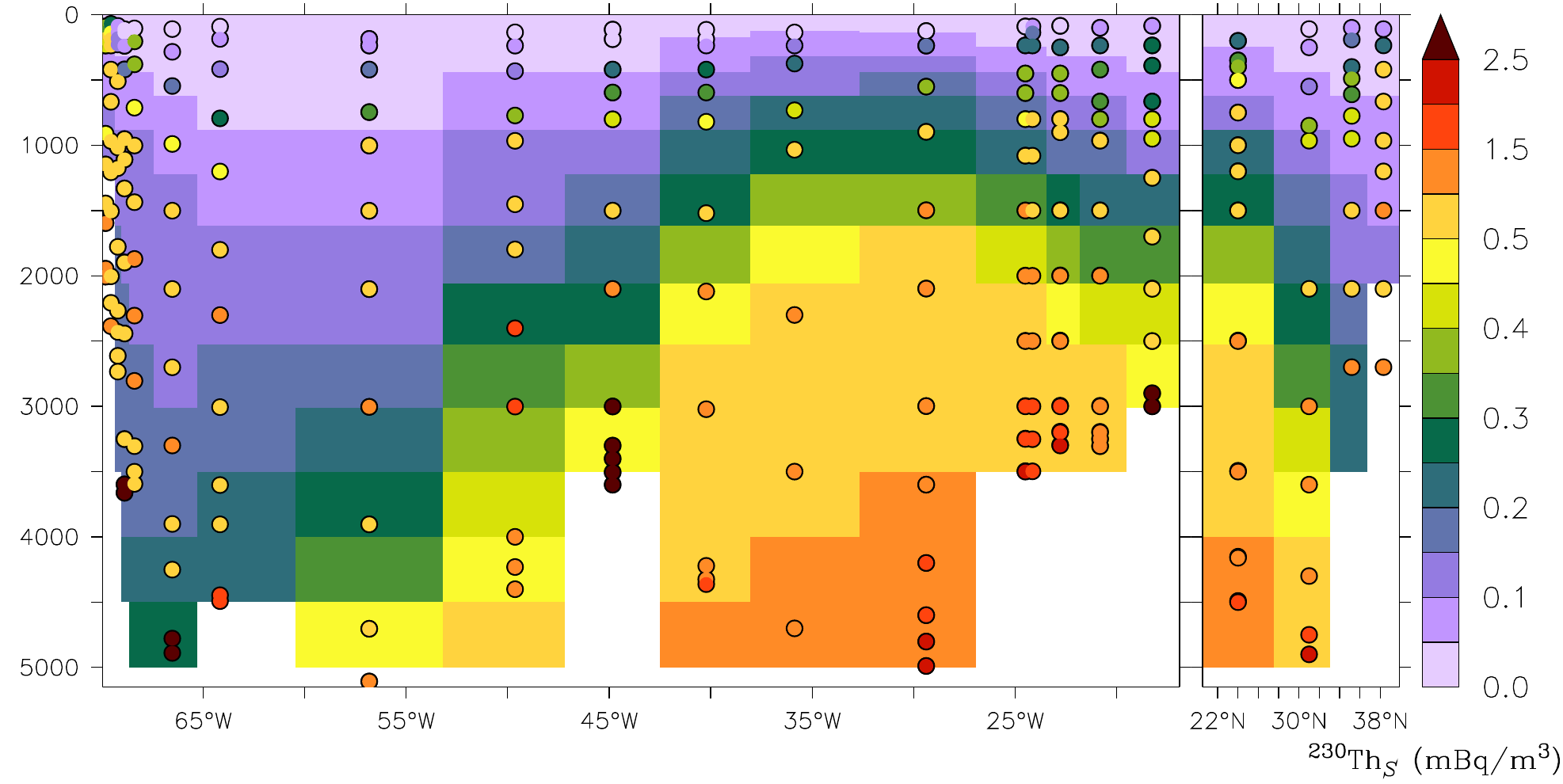}
        \label{fig:Th230s_GA03}
    }
    \subfloat[\chem{[^{231}Pa]}$_S$]{
        \includegraphics[width=.5\linewidth]{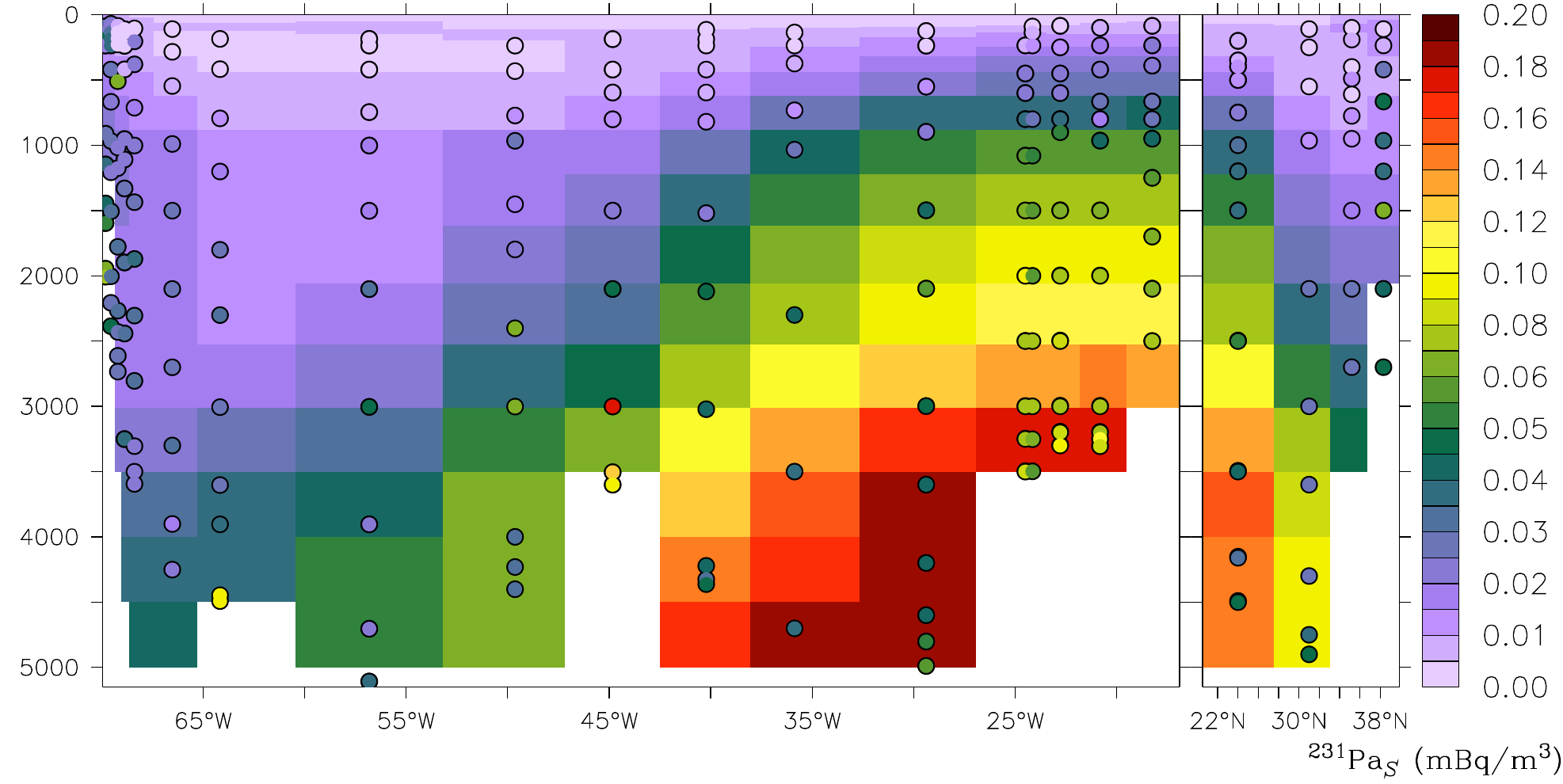}
        \label{fig:Pa231s_GA03}
    }\\
    \subfloat[\chem{[^{230}Th]}$_B$]{
        \includegraphics[width=.5\linewidth]{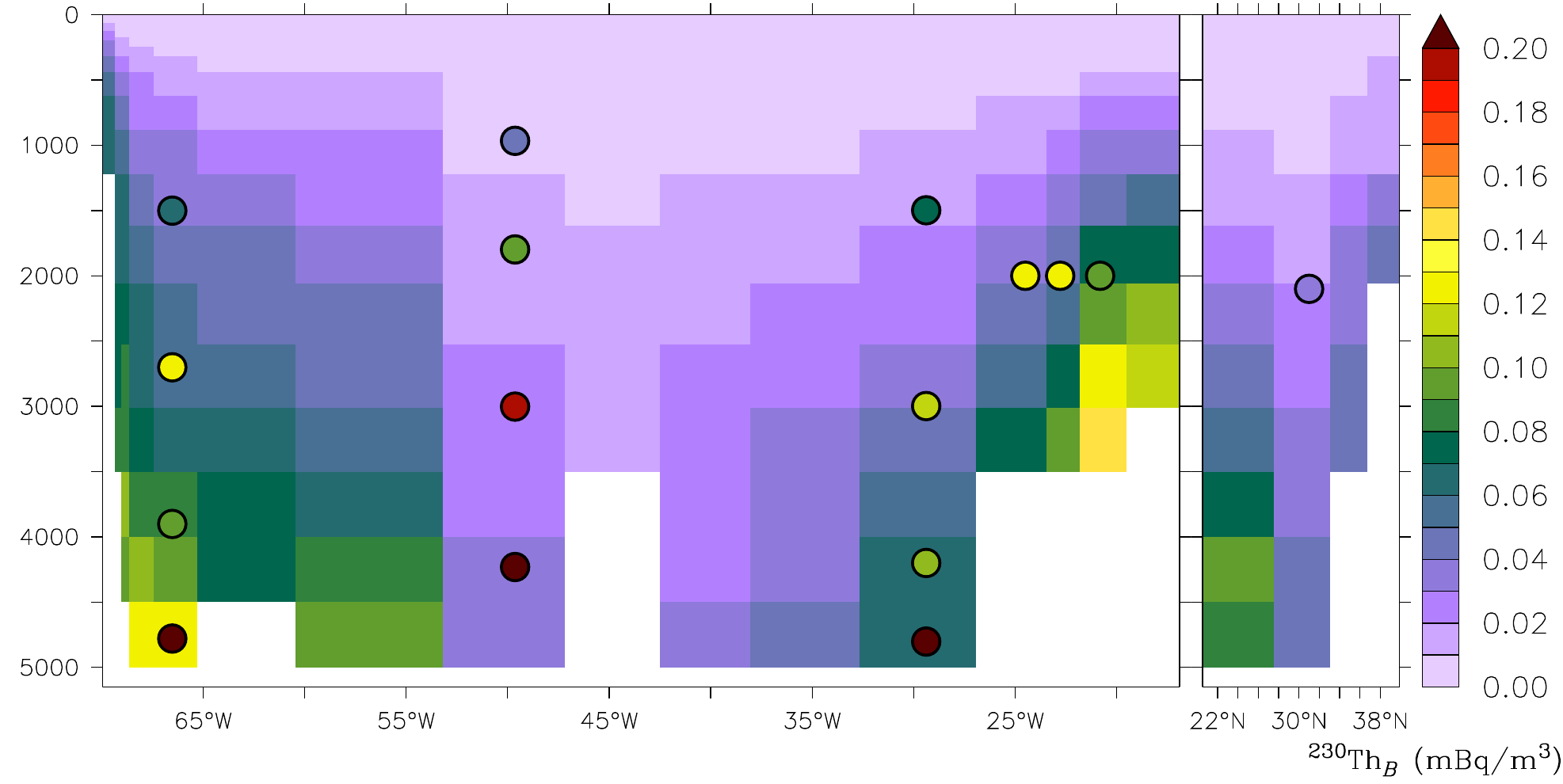}
        \label{fig:Th230f_GA03}
    }
    \subfloat[\chem{[^{231}Pa]}$_B$]{
        \includegraphics[width=.5\linewidth]{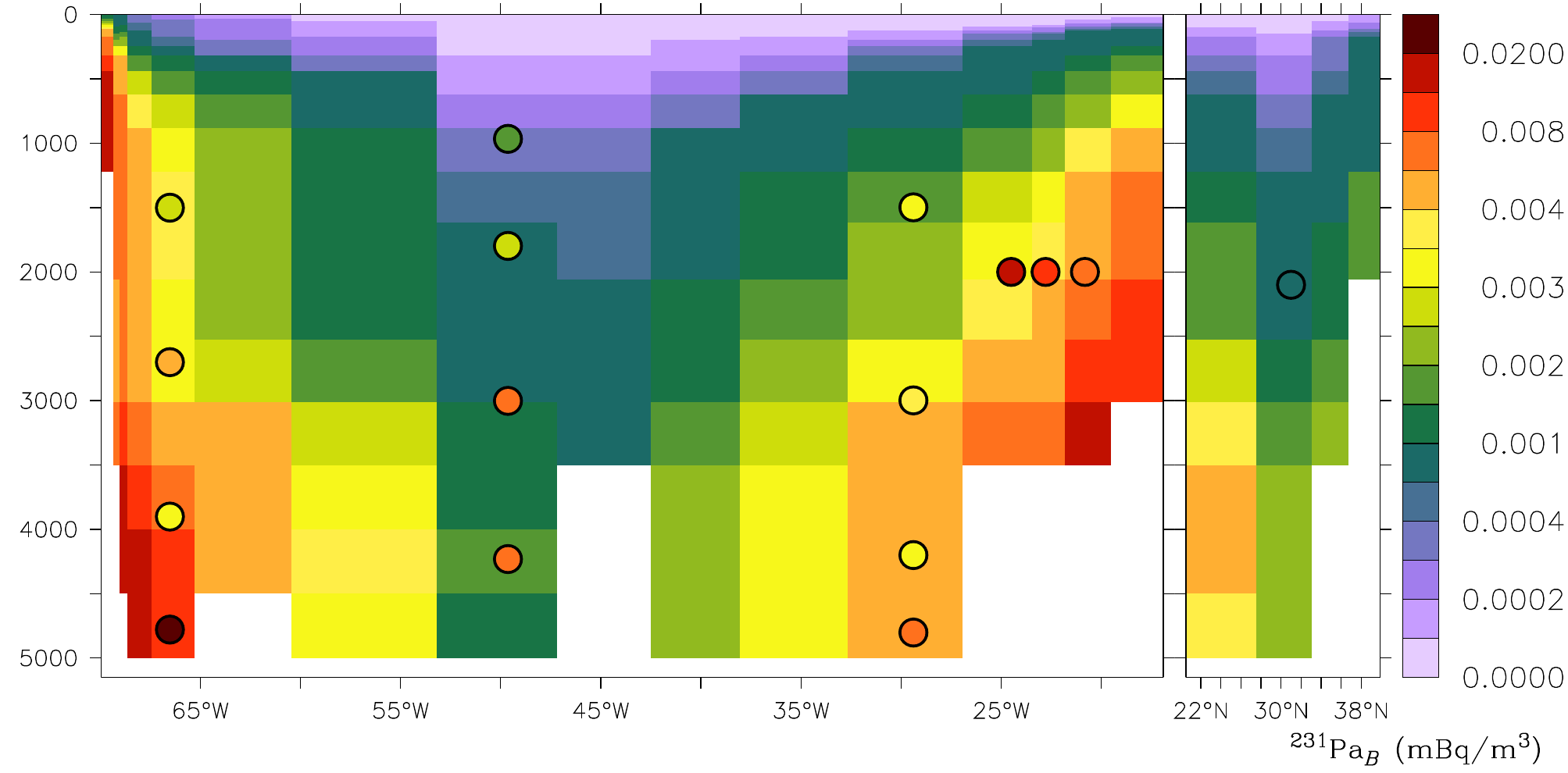}
        \label{fig:Pa231f_GA03}
    }
    \caption{Radionuclide concentrations at the GA03 North Atlantic \textsc{Geotraces} transect.
(a) and (b) display dissolved concentrations, (c) and (d) show the amounts on
small particles, and (e) and (f) those on large particles.
Everything is in (\si{\milli\becquerel\per\cubic\metre}).
Observations \citep{hayes2015:thpa} are shown as coloured dots.}
    \label{fig:radio_GA03}
\end{figure*}

\chem{[^{230}Th]}$_D$ stands for the concentration of \chem{^{230}Th} that is in the dissolved phases.
Similarly, \chem{[^{230}Th]}$_S$ is the concentration of \chem{^{230}Th} that is adsorbed onto the small
particles $S$, \chem{[^{230}Th]}$_B$ the concentration adsorbed onto the large particles $B$; and
similarly for~\chem{^{231}Pa}.
The modelled dissolved distributions of \chem{^{230}Th} and \chem{^{231}Pa} are underestimated at the
surface (Figs~\ref{fig:Th230d_4depths}
and~\ref{fig:Pa231d_4depths}).
Below the surface, \chem{[^{230}Th]}$_D$ is much better captured by the model (also
Fig.~\ref{fig:Th230d_GA03} for the GA03 transect).
However, observations show lower values near the bottom in the western part of
the GA03 section and at 45\degree\,W, associated with more intense scavenging
related to respectively nepheloid layers and manganese oxides from hydrothermal
vents, which are not produced in the simulation (Figs~\ref{fig:Th230d_GA03}
and~\ref{fig:Pa231d_GA03}).

In the intermediate and deep waters of all the oceans, modelled \chem{[^{231}Pa]}$_D$ is
generally of the correct order of magnitude but is strongly overestimated below
2500\;m depth (Fig.~\ref{fig:Pa231d_GA03}), especially in the Pacific Ocean
(Fig.~\ref{fig:Pa231d_4depths}c,d).


The concentrations of the adsorbed phases of \chem{^{230}Th} at GA03 are
presented in Figs~\ref{fig:Th230s_GA03} and~\ref{fig:Th230f_GA03}.
In the deep ocean, modelled \chem{[^{230}Th]}$_S$ and \chem{[^{230}Th]}$_B$ have
lower values than \chem{[^{230}Th]}$_D$, which is consistent with observations.
Furthermore, compared with \citet{dutay2009}, we have notably improved
\chem{[^{230}Th]}$_S$ and \chem{[^{230}Th]}$_B$.
Other global modelling studies have not reported adsorbed concentrations of
\chem{[^{230}Th]}$_D$ and \chem{[^{231}Pa]}$_D$.
However, the model still underestimates the observed \chem{[^{230}Th]}$_S$ and
\chem{[^{230}Th]}$_B$.

Contrarily, small \chem{^{231}Pa} particles are overestimated in our model
below 1500\;m
at the GA03 transect (Fig.~\ref{fig:Pa231s_GA03}).
Large \chem{^{231}Pa} particles are simulated more realistically (Fig.~\ref{fig:Pa231f_GA03}).

\begin{figure*}
\centering
\includegraphics[width=.49\linewidth]{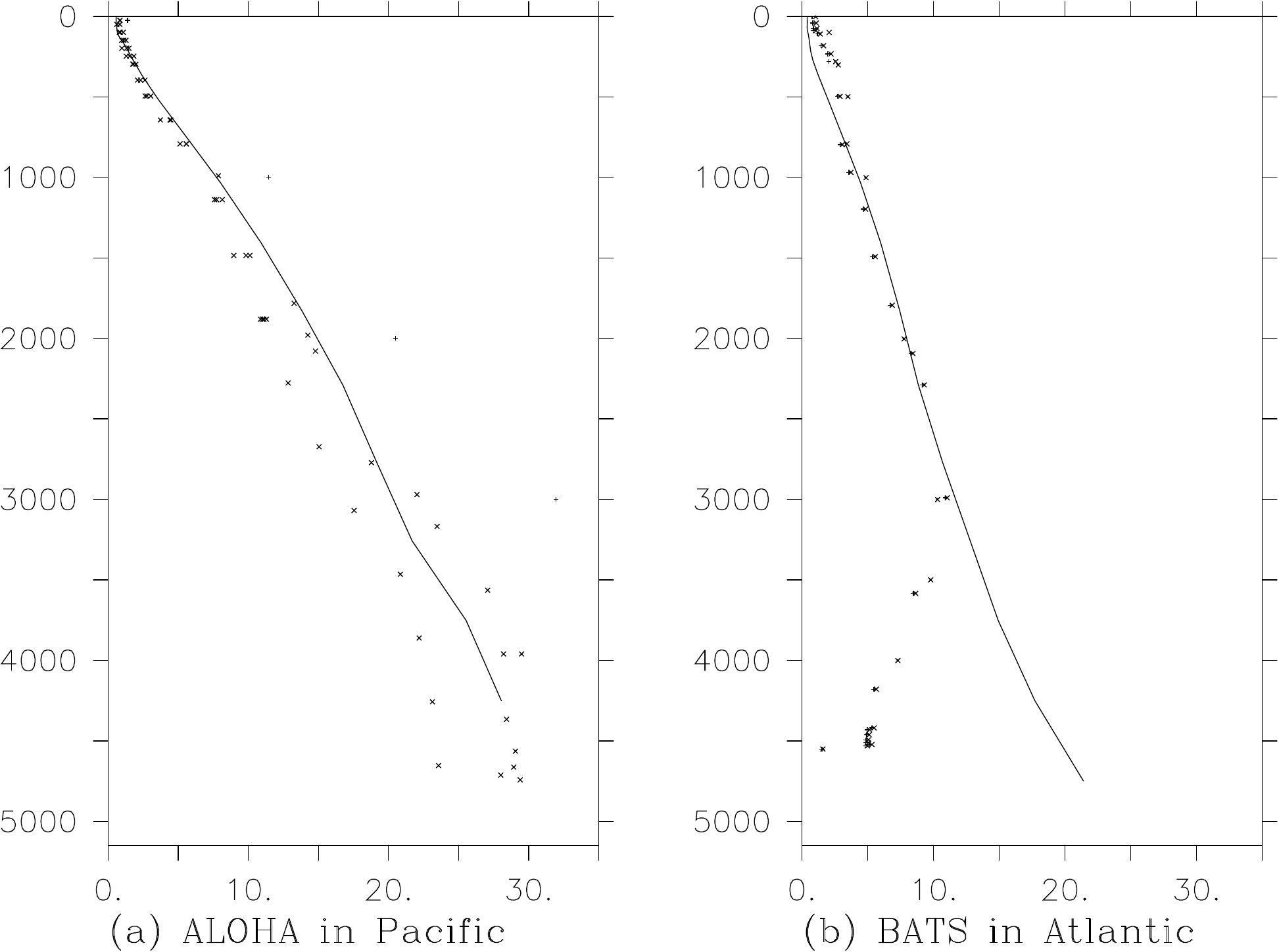}
\includegraphics[width=.49\linewidth]{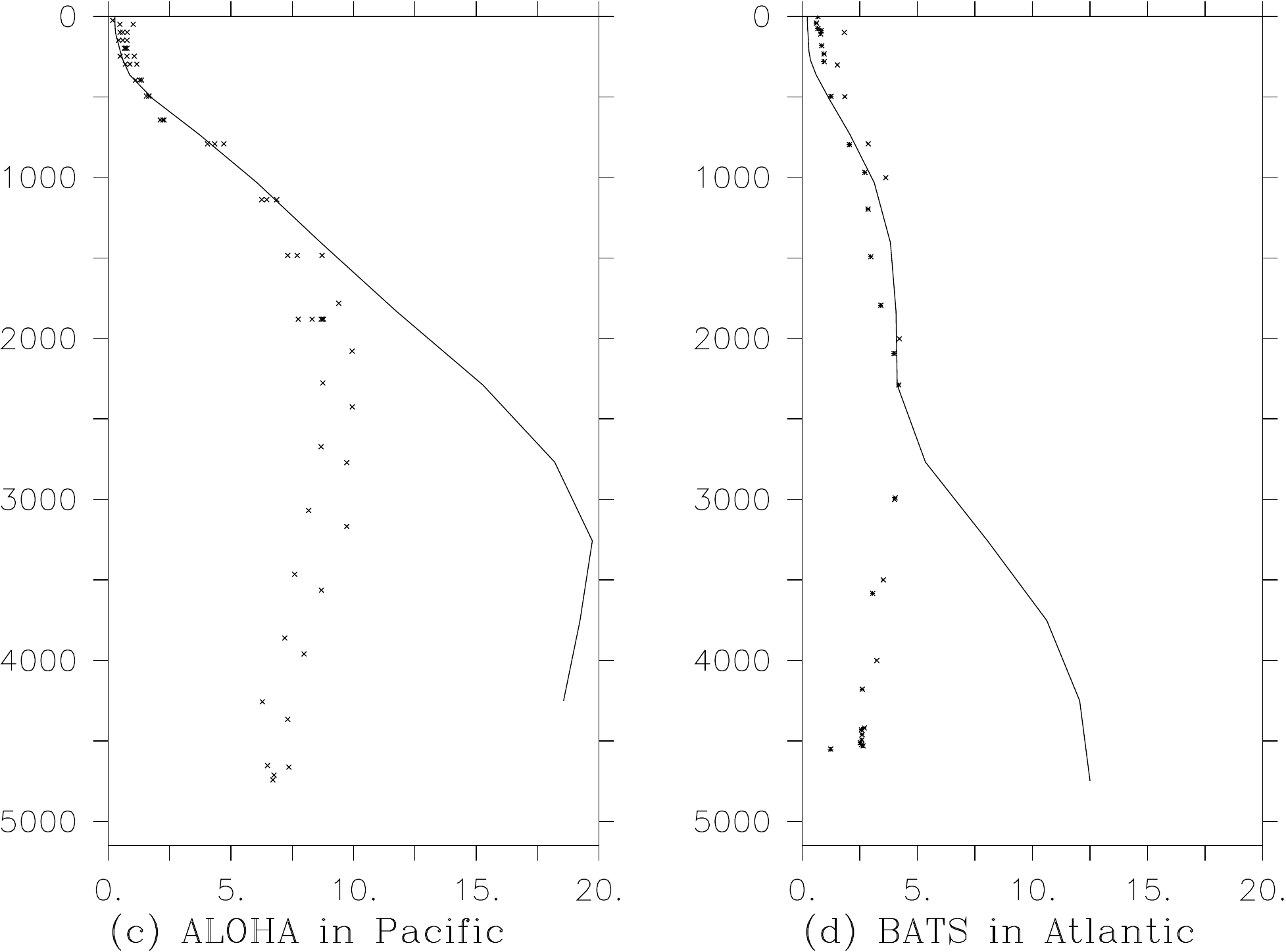}
\caption{Concentrations at stations in the North Pacific (around ALOHA station,
158\degree\,W, 23\degree\,N) and North Atlantic Oceans (around the BATS station,
64\degree\,W, 29\degree\,N).
The left two panels (a,b) display \chem{[^{230}Th]}$_D$, and the right two (c,d) \chem{[^{231}Pa]}$_D$.
The line is the model, and the speckles are the observations.}
\label{fig:profiles}
\end{figure*}

To show clearly the change of \chem{[^{230}Th]}$_D$ and \chem{[^{231}Pa]}$_D$ with respect with depth, profiles
are plotted in Fig.~\ref{fig:profiles}.
\chem{[^{230}Th]}$_D$ is linearly increasing downwards, which matches with the observations, but
below 3000\;m near Bermuda, \chem{[^{230}Th]}$_D$ increases where it should decrease according to
the observations.
The shape of \chem{[^{231}Pa]}$_D$ matches that of the observations for the upper 2\;km, below
which our model overestimates the observations up to a factor four at the
bottom of the ocean.

Globally, only 0.3\,\% of the \chem{^{231}Pa} sits in the particulate pool, and
for \chem{^{230}Th} this is 1.3\,\%.
The rest is in the dissolved phase.
Observational data shows that more than 10\,\% of \chem{^{230}Th} sits in the particulate
pool, typically adsorbed onto small calcium carbonate particles.
This can be explained by the fact that we only have big, fast-sinking
\chem{CaCO_3} particles in the model.
Since the big particles are removed quickly, these cannot act as a realistic
\chem{CaCO_3}-associated \chem{^{230}Th} pool without depleting \chem{^{230}Th}, and thus we
underestimate the particulate radionuclide concentrations.
Table~\ref{tab:contrib} gives two types of information on the adsorbed phases.
One is the fraction of radionuclide carried by each phase (``global stock'').
The other is the fraction of the flux carried by each type of particle (the
other rows) for the global ocean and two regions of the ocean.
The stock and the flux are not necessarily the same, because the big particles
have a different settling velocity from the small particles.

\begin{table}
\centering
\begin{tabular}{ll|rrrr}
\toprule
\multicolumn{2}{l}{\%}      & POC   & \chem{bSiO_2} & \chem{CaCO_3} & Litho. \\
\midrule
\multirow{4}{*}{\chem{^{231}Pa}}
    & global stock          & 35.5  & 10.5          &  3.1          & 50.8  \\
    & global particle flx    & 16.5  & 53.5          & 15.8          & 14.2  \\
    & North Atlantic flx    & 16.5  & 25.7          & 15.6          & 42.1  \\
    & Southern Ocean flx    &  5.6  & 93.7          &  0.6          &  0.1  \\
\midrule
\multirow{4}{*}{\chem{^{230}Th}}
    & global stock          & 21.3  &  3.2          & 24.3          & 51.2  \\
    & global particle flx    &  5.9  &  9.9          & 75.8          &  8.7  \\
    & North Atlantic flx    &  4.4  &  2.8          & 69.9          & 22.9  \\
    & Southern Ocean flx    &  9.7  & 71.1          & 18.9          &  0.3  \\
\bottomrule
\end{tabular}
\caption{Relative global budget of \chem{^{231}Pa} and \chem{^{230}Th} in the different particulate
phases (``global amount''), globally averaged fluxes in the different phases
(``global flux'') and the same but for the North Atlantic Ocean
(44--24\degree\,W, 25--56\degree\,N) and the Weddell Sea (44--24\degree\,W,
76--63\degree\,S).}
\label{tab:contrib}
\end{table}

Only a small amount of \chem{^{230}Th} is adsorbed onto biogenic silica (3\,\%), though
the amount is larger for \chem{^{231}Pa} (11\,\%).
However, the global settling flux of \chem{^{231}Pa} is primarily (54\,\%) determined by
\chem{bSiO_2}, particularly due to the high \chem{bSiO_2} fluxes in the Southern
Ocean.
Most of the modelled particulate \chem{^{231}Pa} and \chem{^{230}Th} is on the
lithogenic particles, but lithogenic particles account for only 14\,\% of the
\chem{^{231}Pa} flux and 9\,\% of the \chem{^{230}Th} flux.
If lithogenic particles from nepheloid layers were included, this fraction
would be higher.
The flux contribution is also different from what is in the pools when we look
at \chem{CaCO_3}: only 24\,\% of the \chem{^{230}Th} is on calcium carbonate, but
\chem{CaCO_3} is for 76\,\% responsible for the \chem{^{230}Th} export.
These discrepecies between the contribution of each particle type to the flux
and to the stock of radionuclides arise from the different speeds for each type
of particle.

The relative contributions of the different particulate phases also vary for
different regions of the ocean.
Opal is for 94\,\% responsible for \chem{^{231}Pa} export in the Southern Ocean.

Compared to \citet{dutay2009}, the improvement of the \chem{^{230}Th} profile
shape is only to a small extent due to the large improvement in the POC
representation \citep{aumont2017}, because the POC ensures only 6\,\% of the
vertical transport of \chem{^{230}Th} (Table~\ref{tab:contrib}).
The improvement is mostly due to the addition of lithogenic dust
particles and the adjusted \chem{CaCO_3} dissolution.
Lithogenic particles and \chem{CaCO_3} hardly dissolve compared to POC and hence
the downward particle flux remains constant, which is in agreement with 1-D
models which show linear profiles of \chem{^{230}Th}$_D$ \citep[e.g.][]{roybarman2009}.

\subsection{Sedimentation flux ratio of \chem{^{231}Pa}/\chem{^{230}Th}}

\begin{figure}
\centering
\includegraphics[width=\linewidth]{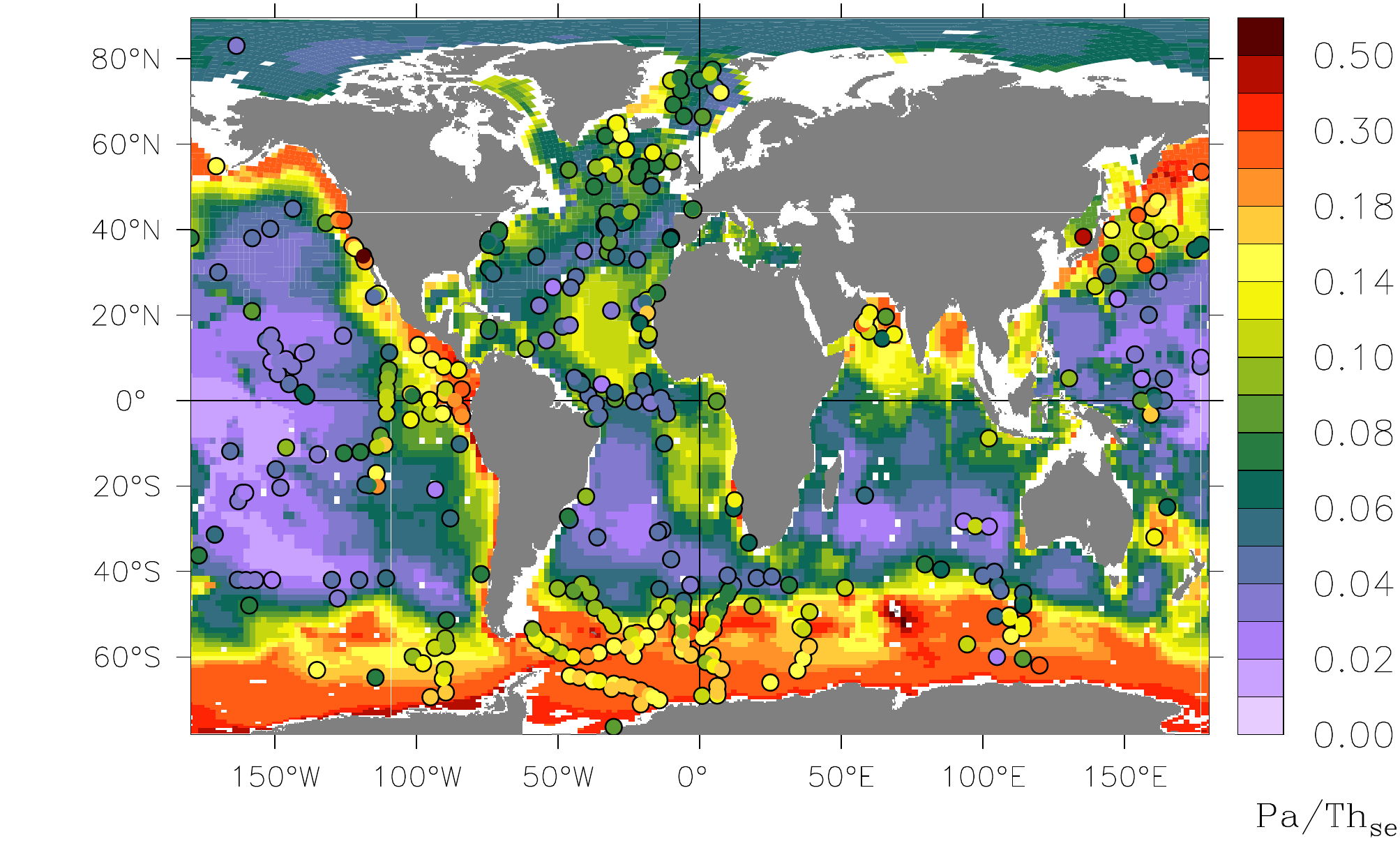}
\caption{Sedimented \chem{^{231}Pa}/\chem{^{230}Th} ratio.
        Top core measurements are presented as dots.
        This is the standard simulation with $K_{\chem{Pa},\chem{bSiO_2}} /
        K_{\chem{Th},\chem{bSiO_2}} = 1$.}
\label{fig:sed}
\end{figure}

The sedimentation flux of the total adsorbed \chem{^{231}Pa}/\chem{^{230}Th}
ratio captures the general patterns of the upper sediment core
\chem{^{231}Pa}/\chem{^{230}Th} concentration ratio (Fig.~\ref{fig:sed}).
Consistent with observations, the model produces more elevated
\chem{^{231}Pa}/\chem{^{230}Th} ratio values in the Southern Ocean, the northern
part of the Indian Ocean, the Pacific Ocean, and along the coastal upwelling
regions.
However, at some places, like much of the Southern Ocean, our model tends to
overestimate the ratio derived from the sediment core observations.

\section{Discussion}                    \label{sec:discussion}

We succeed to generate the global patterns of the dissolved nuclide
concentrations.
However, we underestimate radioactivities in the surface waters
(Figs~\ref{fig:Pa231d_4depths}a and~\ref{fig:Th230d_4depths}a).
This shortcoming may be due to the instant equilibration between the dissolved
and the adsorbed phases \citep{henderson1999}.
After decay of \chem{U} into \chem{^{231}Pa} and \chem{^{230}Th}, the tracers
instantly adsorb onto the particles in the surface ocean and are exported from
the mixed layer.
For future model developments, it is of interest to test if a non-instant
equilibration model (parameterised with adsorption and desorption coefficients)
yield better surface values.

The reversible scavenging model uses partition coefficients $K_{ij}$ for
the different isotopes $i$ and particles $j$.
Many studies provide constraints on these coefficients but
estimations vary strongly between different studies.
Small particles are the most important scavengers and largely control the shape
of the vertical profile of the tracers \citep{dutay2009}.
This is explained by the specific surface of small particles, which is larger
than that of big particles.
Therefore higher adsorption values are set for small particles compared to
the bigger particles of the same type (namely, 5 times higher for POC, and 10
times higher for lithogenic particles).
Compared to \citet{dutay2009}, we succeed to simulate the tracer
concentrations using reasonable $K$ values.
This is due to improvement in (especially small) particle concentrations, which were
generated with the lability parametrisation of POC \citep{aumont2017}, the
change of the \chem{CaCO_3} dissolution parameterisation, and the addition of
lithogenic dust in our ocean model.
This shows that at least two particle size classes are needed to simulate
dissolved and particulate thorium and protactinium.
Our model does not include small particles for all types (there are
neither small \chem{CaCO_3} nor small \chem{bSiO_2} particles), even though we
expect that better results can be reached upon adding those particles akin to
small POC and small lithogenic particles.

\begin{figure*}
\centering
\includegraphics[width=.49\linewidth]{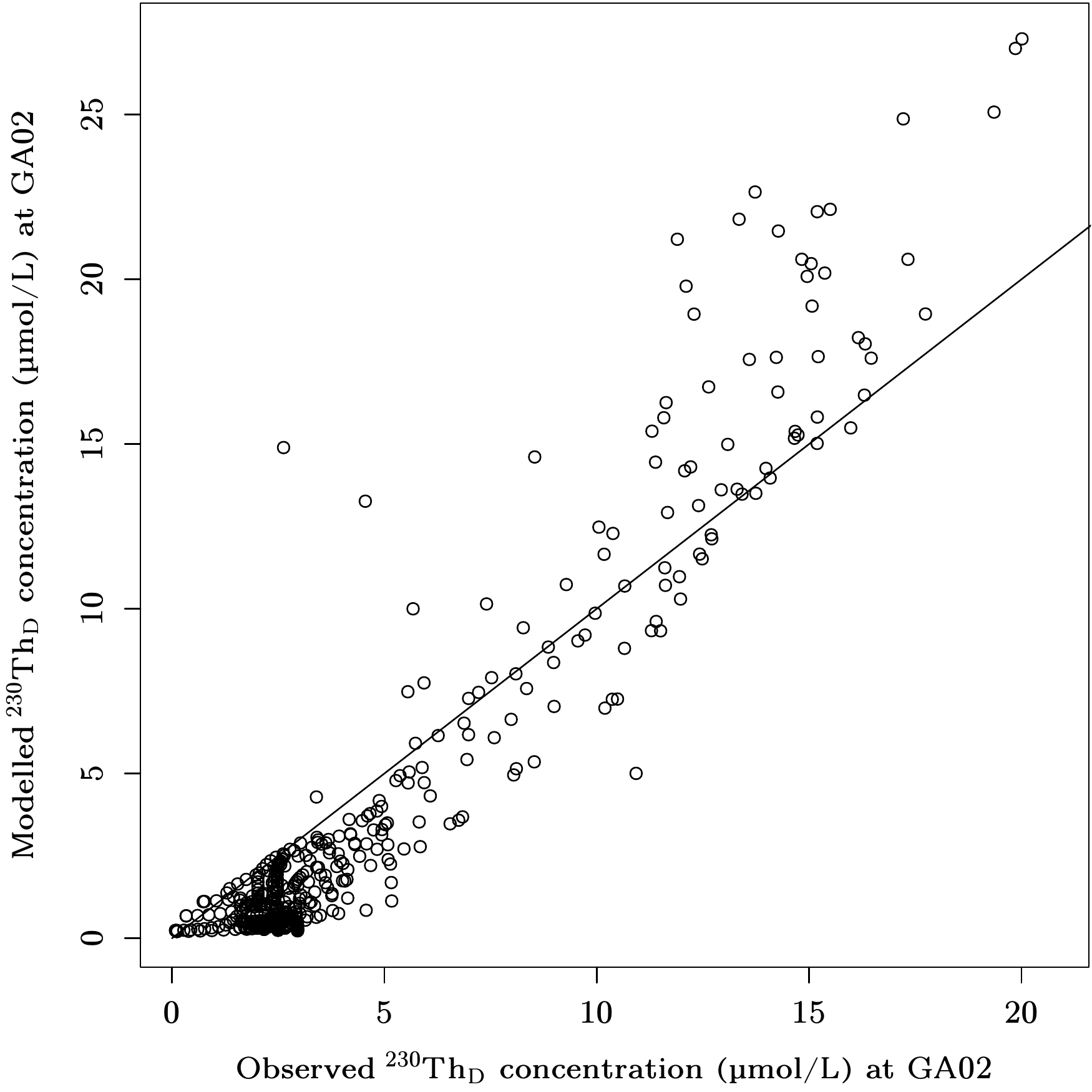}
\includegraphics[width=.49\linewidth]{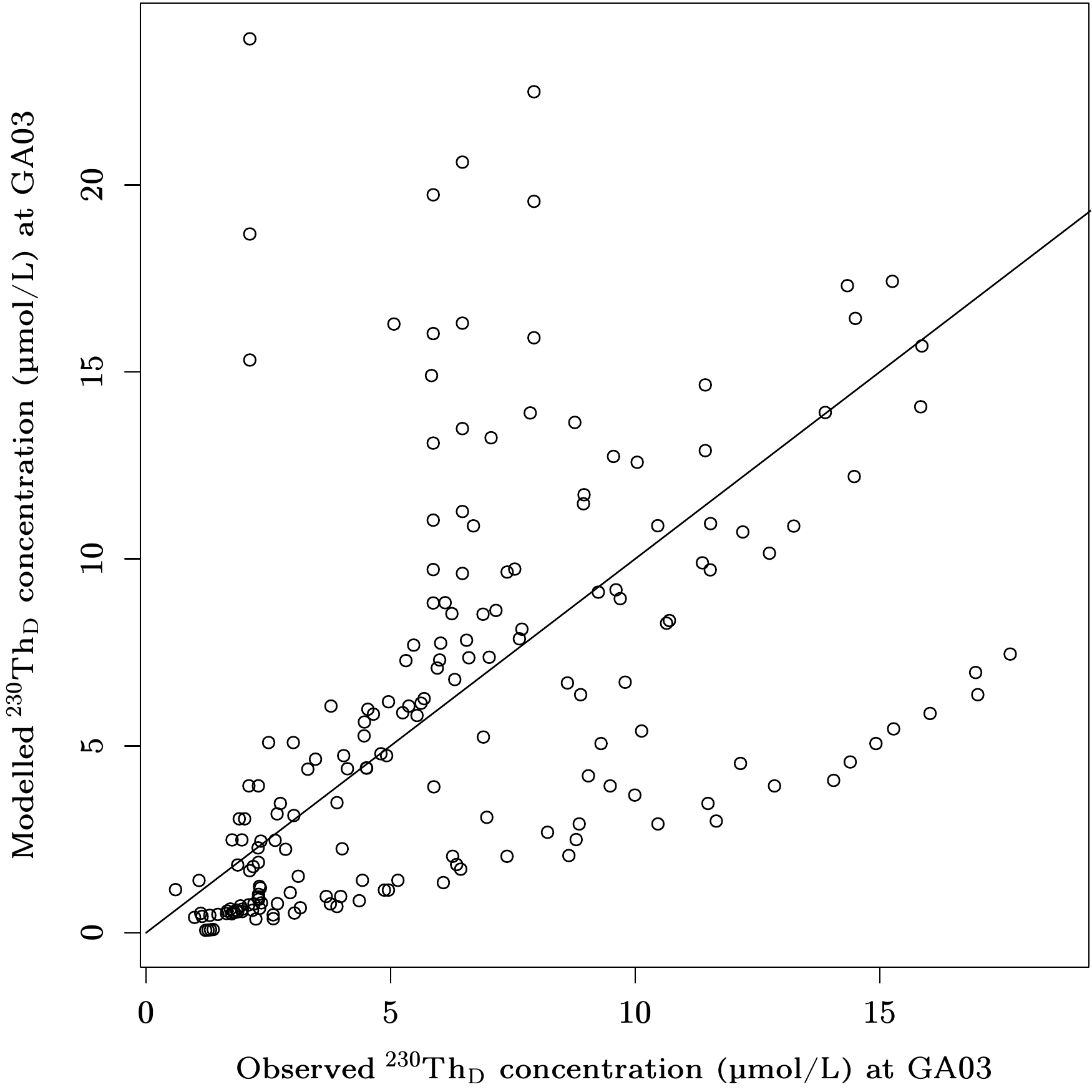}
\caption{Modelled ($y$-axis) versus observed ($x$-axis) dissolved
\chem{^{230}Th} concentration for the West Atlantic GA02 transect (left panel)
and the North Atlantic GA03 transect (right panel).
The units are \si{\milli\becquerel\per\cubic\metre}.
The line is $y = x$.}
\label{fig:Th_scatter}
\end{figure*}

On the GA02 transect, the correlation coefficient $r$ of the data--model comparison
of \chem{[^{230}Th]}$_D$ is 0.83, which is the same as reported by
\citet{rempfer2017} in their simulations without bottom-boundary scavenging.
However, on the GA03 section, $r = 0.21$, which may be low due to several reasons.
The correlation between model and observation are presented as scatter plots
(Fig.~\ref{fig:Th_scatter}).

Firstly, the hydrothermal plume over the ridge results in \chem{^{230}Th}-poor seawater that
is not reproduced by the model.
Secondly, in the western boundary currents, very low dissolved \chem{^{230}Th}
concentrations are observed due to strong nepheloids and/or possibly strong deep
ventilation from the north.
Neither of these processes are represented in the model.
Missing the nepheloids has an especially large impact on the GA02 transect,
which has a relatively high amount of measured profiles near the western
boundary.
Finally, between 32 and 20\degree\,W, the model tends to underestimate
\chem{^{230}Th}.
The value of the correlation index is not high, but the goal of this study is
not to give a perfect model--data comparison.

The provided dust flux does not distinguish between different dust particle
sizes or types.
Futhermore, dust deposition is uncertain; therefore we chose a dust flux that
was available to the ORCA2 configuration of NEMO, and its biogeochemistry has
been tested with that dust flux.
However, given the significant impact of dust particles to thorium and
protactinium concentrations, it would be useful to look into the effect of
different dust deposition fields, and of using different dust particles size
classes.
A good model of \chem{^{230}Th} (and other isotopes of thorium) could even
constrain the sporadic and uncertain dust deposition, but all this falls outside
the scope of the present study.

Our model adds 80\,\% of the dust deposition into the large lithogenic
particle compartment ($\varnothing > \SI{50}{\micro\metre}$), and only 20\,\% goes
into the small compartment ($\varnothing \leq \SI{50}{\micro\metre}$).
This is necessary to reproduce the distributions of lithogenic particles.
However, airborne dust particles typically have a diameter of smaller than \SI{20}{\micro\metre}
\citep{book::knippertz2014} and thus fall clearly within the small size class of
less than \SI{50}{\micro\metre}.
Recently, there have been observations of aerosols larger than
\SI{20}{\micro\metre} \citep{does2016}, but they do not reach far enough into
the atmosphere above the open ocean to explain the high concentration of large
lithogenic ocean particles.
The explanation for the apparently required high fraction of large lithogenic
particles is that smaller aerosols aggregate in the upper layers of the ocean.
Moreover, though more hypothetically, there may be strong aggregation with
biogenic particles just below the surface, below which the aggregates partly
disaggregate again to result in the $\sim$0.2 big lithogenic fraction in the
deep ocean (Fig.~\ref{fig:fLith_g_GA03}).

The underestimation of lithogenic particle concentrations at the western
boundary is expected.
The reason is that we only have dust deposition as a source of lithogenic
particles, whereas nepheloid layers are not included.
Nepheloid layers are at least locally an important source of lithogenic (and
biogenic) particles \citep{lam2015:size}.
With the transport of \chem{^{230}Th} and \chem{^{231}Pa} through the western
boundary current, a significant portion may be scavenged.
Moreover, the periodic transport through the North Atlantic Gyre would result in
lower \chem{[^{230}Th]}$_D$ and \chem{[^{231}Pa]}$_D$ throughout a large part of
the North Atlantic Ocean, possibly improving the concentrations in the deep
ocean.
Therefore it will be useful to include nepheloid layers in the future.
We have not done this so far, because we do not know how to model nepheloid
layers.
Except for trivial models, like the one of \citet{rempfer2017} who forced an
additional constant scavenging rate in the bottom box of the ocean model, no
large-scale, prognostic nepheloid models have been developed.


We overestimate the radionuclide activity in the deep ocean, which is partly
because we underestimate particle concentrations.
Especially thorium adsorbs well onto lithogenic particles, which are
underestimated in the West Atlantic Ocean (in the subsurface, west of about
43\degree\,W\@, Fig.~\ref{fig:Lith_t_GA03}).
Below 2\;km depth, also POC is underestimated (Fig.~\ref{fig:part_GA03}a), even though the lability
parameterisation improved the distribution by over an order of magnitude
compared with previous versions of \textsc{Pisces} \citep{aumont2017}.
Small POC and the small lithogenic particles are the only small
particles.
Their underestimation in the deep ocean results in an overestimation of the
radionuclides in the deep ocean.
In our simulation, \chem{[^{230}Th]}$_D$ is not that much overestimated (outside the Arctic Ocean).
However, \chem{[^{231}Pa]}$_D$ is strongly overestimated in the deep Pacific and Atlantic Oceans.

\begin{figure}
\centering
\includegraphics[width=\linewidth]{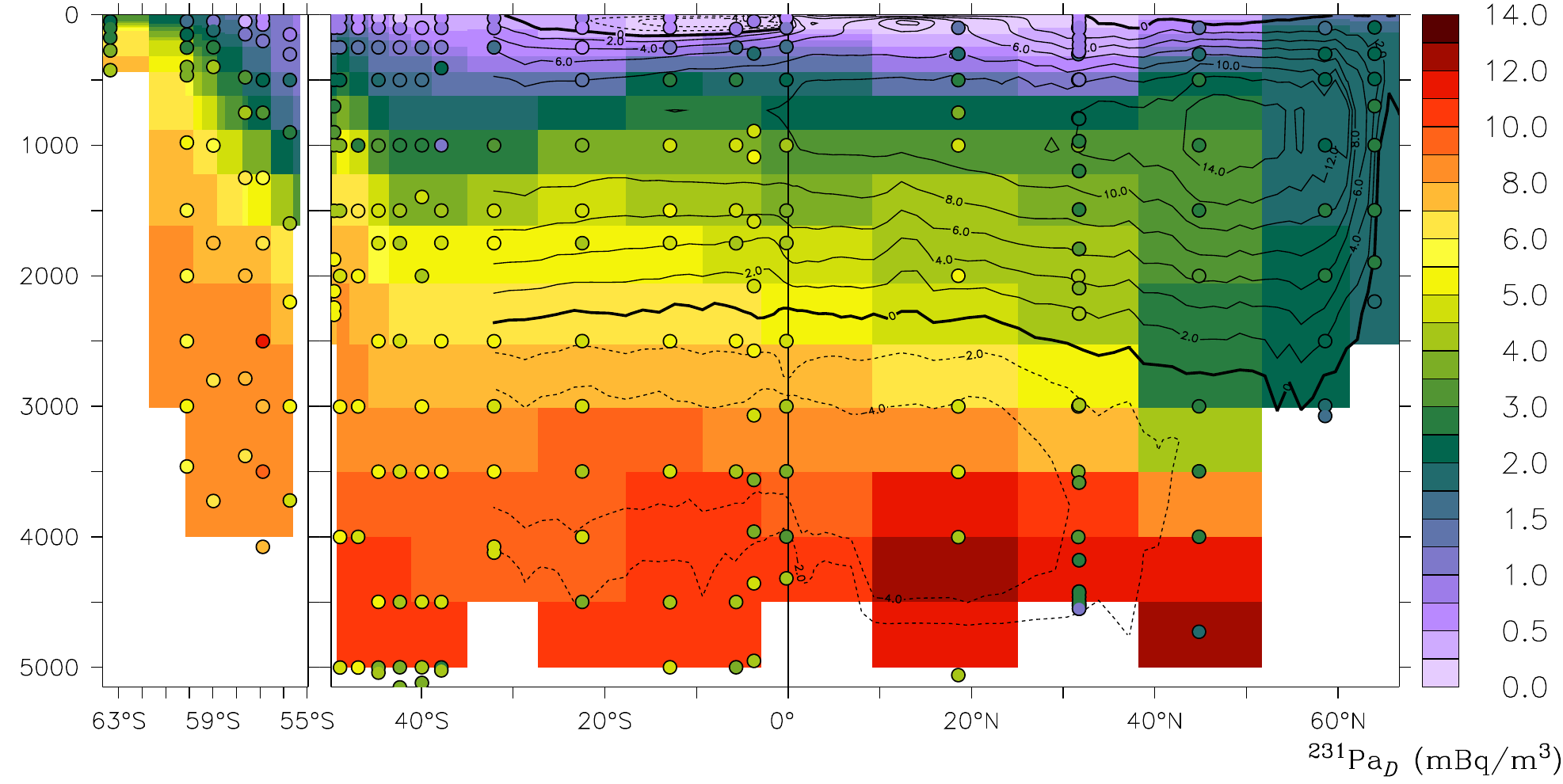}\\
\includegraphics[width=\linewidth]{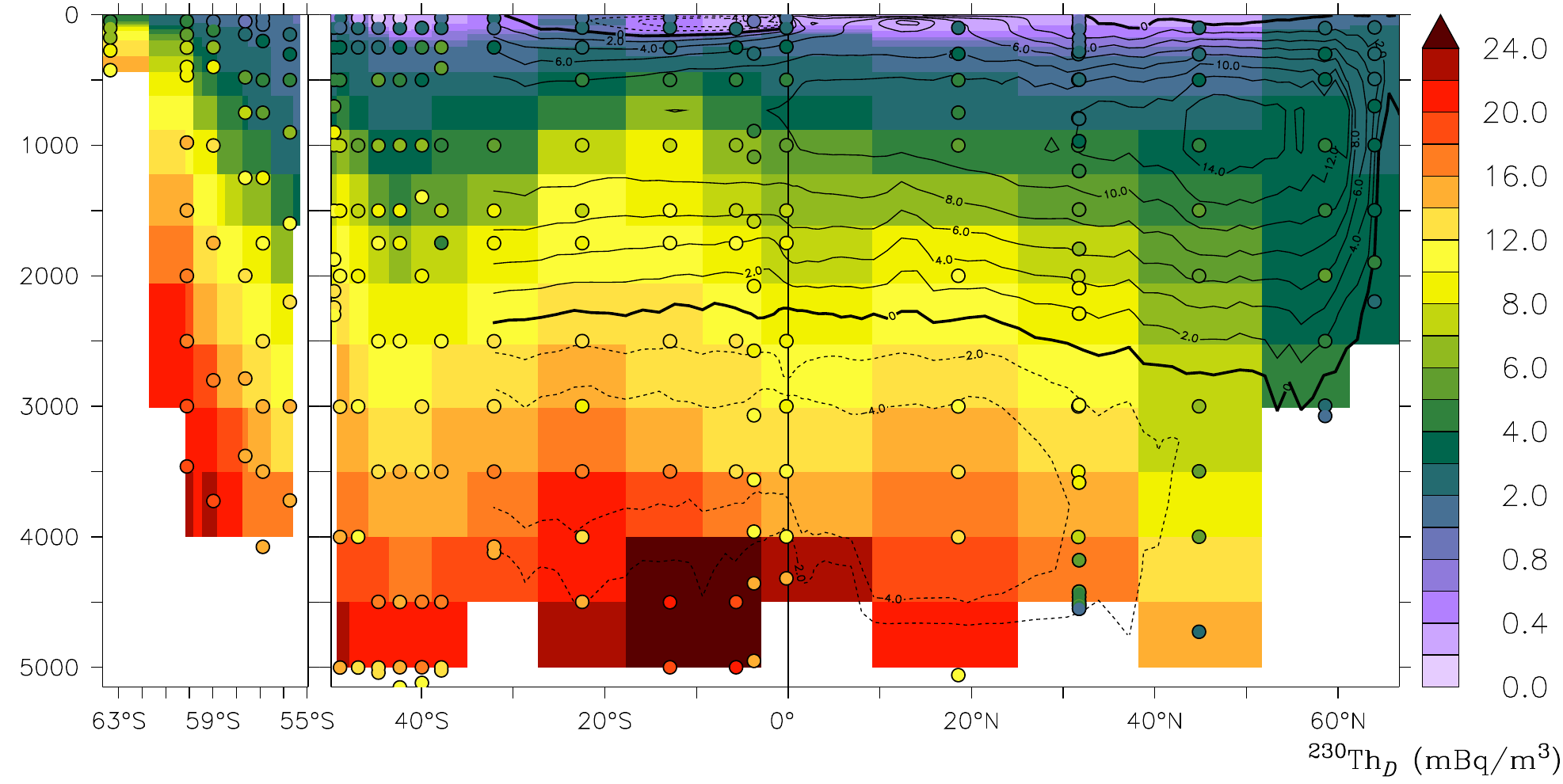}
\caption{Dissolved protactinium-231 (upper panel), and thorium-230 (lower panel)
        at the Drake Passage (GIPY5) and West Atlantic (GA02) transects
        (\si{\milli\becquerel\per\cubic\metre}); observations as dots.
        The Atlantic OSF, in Sv, is superimposed as a contour.}
\label{fig:GA02}
\end{figure}

Clearly the model does not remove \chem{^{231}Pa} efficiently enough from the
deep ocean (Figs~\ref{fig:Pa231d_4depths}c,d, \ref{fig:Pa231d_GA03} and
Fig.~\ref{fig:GA02}--upper panel).
Two likely reasons may be that the waters are not well ventilated in the
model (older than in reality), or that there is not enough scavenging.
The Atlantic OSF (Fig.~\ref{fig:osf}) shows that the upper overturning cell is
too shallow compared to observational studies, so there is too little
ventilation.
This means that \chem{^{231}Pa} can build up in the deep water due to sluggish
ventilation of the physical model \citep{dutay2002,biastoch2008:MOC}.
Moreover, the AABW has weakened in the deep North Atlantic Ocean near the
high-\chem{[^{231}Pa]}$_D$ region (Fig.~\ref{fig:GA02}--upper panel), which may
also contribute to the high dissolved \chem{^{231}Pa} concentrations.
The volume transport of the lower cell is about 6\;Sv near 20\degree\,N and
still 2\;Sv near 40\degree\,N, which is in the order of magnitude of what is
reported by literature.
However, it is not obvious whether the weak overturning is large enough to explain the
discrepancy between the modelled and the observed \chem{[^{231}Pa]}$_D$.
In reality, there are many regions in the North Atlantic Ocean where there is sediment
resuspension and where there are nepheloid layers
\citep{gardner1981,lam2015:size,gardner2017}.
This is consistent with the fact that our model underestimates lithogenic particle
concentrations in the West Atlantic Ocean (Fig.~\ref{fig:Lith_t_GA03}).
The additional lithogenic particles would scavenge more \chem{^{231}Pa}, resulting in lower
\chem{[^{231}Pa]}$_D$.
Even though the enforced scavenging occurs near the floor and the western
boundary of the ocean, a strong flux of water passes through the latter region,
and is transported through the North Atlantic Gyre and through the rest of the
Atlantic, diluting high tracer concentrations.
The small POC and lithogenic particles are now underestimated in much
of the deep ocean.
They would contribute to lower radioactivities as well.

\citet{rempfer2017} confirmed that an additional bottom sink affects
\chem{^{231}Pa} and \chem{^{230}Th} significantly.
They used a homogeneous extra scavenger in their model grid's bottom grid cell.
Therefore, it would be worth testing if a realistic distribution of
nepheloids would result in the right amount of scavenging.

Other particles may include resuspended (nepheloidal) biogenic particles but
also manganese and iron hydroxides that are not included in our model, and a
smaller class of calcium carbonate and biogenic silica particles in addition to
the large-particle classes that are already in the model.
Manganese oxides are available especially near hydrothermal vents, but also
throughout the Pacific Ocean in low concentrations but much higher than in the
Atlantic Ocean.
Indeed, it has been argued that hydrothermal inputs may provide additional
removal of \chem{^{231}Pa} and \chem{^{230}Th} \citep{lopez2015,rutgersvanderloeff2016,german2016}, and recently
this is confirmed based on an analysis of the $\sim$10\degree\,S zonal GP16
transect in the Pacific Ocean \citep{pavia2017}.

\subsection{Effect of scavenging by \chem{bSiO_2} on the \chem{^{231}Pa}/\chem{^{230}Th} sedimentation flux ratio}
\label{sec:opal}

Previously, it was suggested that \chem{Pa} has a stronger affinity for
\chem{bSiO_2} than \chem{Th} has, i.e.\ $F_\chem{Pa/Th,bSiO_2} > 1$
\citep{henderson1999,chase2002,dutay2009}.
This ``standard view'' has weakened over time, so it is often accepted that
$F_\chem{Pa/Th,bSiO_2} \gtrsim 1$ \citep{venchiarutti2011,rutgersvanderloeff2016,rempfer2017}.
We argue here that it is \chem{Th} which has the stronger affinity to
\chem{bSiO_2} ($F_\chem{Pa/Th,bSiO_2} < 1$), similarly (though not quite) like
other particles.

Our standard simulation (with $F_\chem{Pa/Th,bSiO_2} = 1$) yields a too high
sedimentation \chem{^{231}Pa}/\chem{^{230}Th} ratio compared to
top-core sediment observations below diatom-rich areas such as the Southern Ocean.
In that region, \chem{bSiO_2} largely controls the particle scavenging of the two
tracers, especially that of \chem{^{231}Pa} (Table~\ref{tab:contrib}).
In order to estimate how strongly the model depends on the value of the
\chem{bSiO_2} partition coefficient we performed a sensitivity experiment where
we set $K_{\chem{Pa},\chem{bSiO_2}} = \SI{0.4}{\mega\gram\per\gram}$ and
$K_{\chem{Th},\chem{bSiO_2}} = \SI{1.0}{\mega\gram\per\gram}$ such that now
$F_\chem{Pa/Th,bSiO_2} = 0.4$ (Table~\ref{tab:K_d}).

\begin{figure}
\centering
\includegraphics[width=\linewidth]{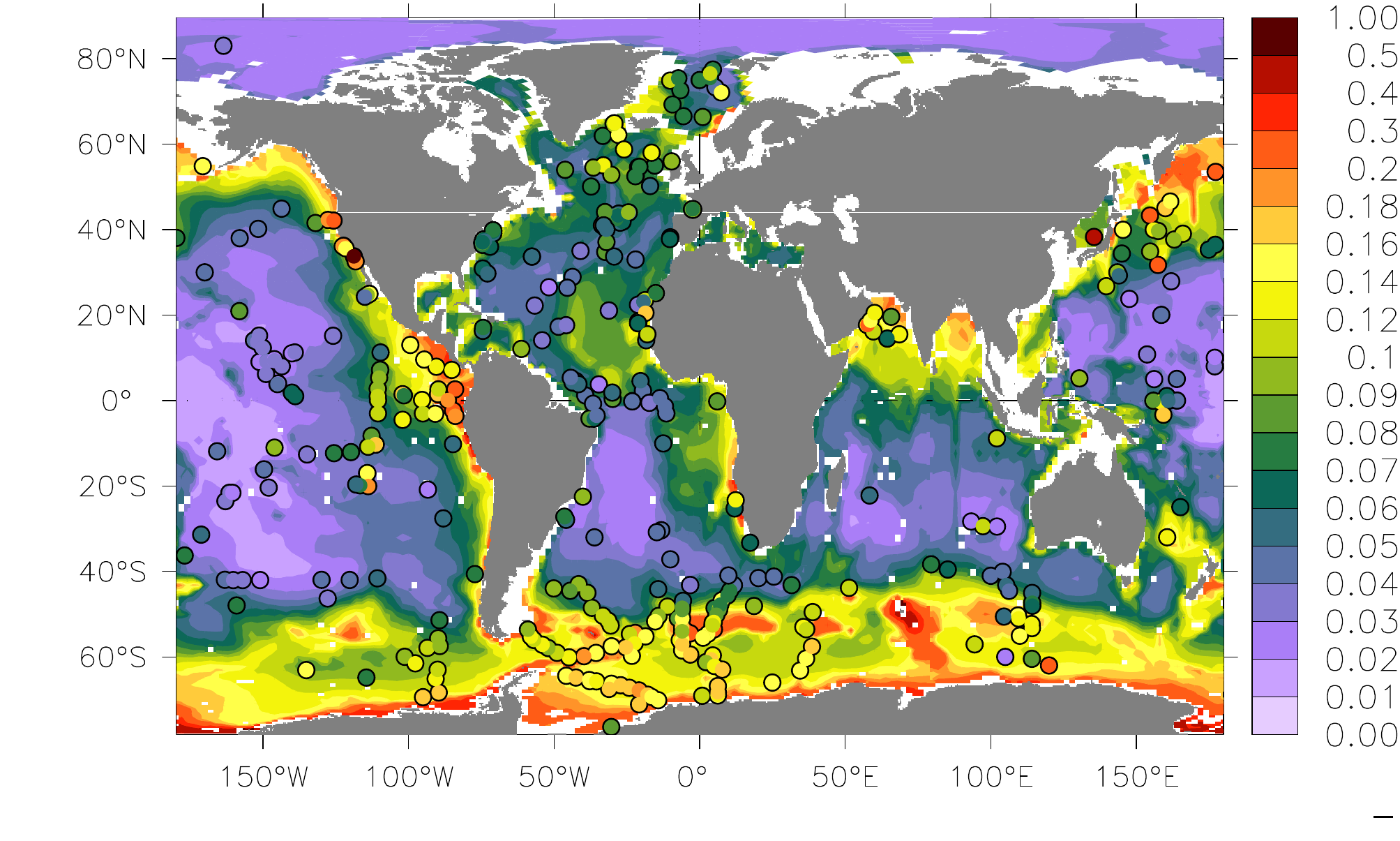}
\caption{Sedimented \chem{^{231}Pa}/\chem{^{230}Th} ratio when we set
$K_{\chem{Pa},\chem{bSiO_2}} / K_{\chem{Th},\chem{bSiO_2}} = 0.4$.
Top core measurements are presented as dots.}
\label{fig:sed_reduced}
\end{figure}

As a result of this reduction, \chem{^{231}Pa}/\chem{^{230}Th} flux ratio is not that much
overestimated anymore in diatom-rich regions (Fig.~\ref{fig:sed_reduced}).
This result suggests that protactinium has a weaker affinity to biogenic silica
than thorium has, though the fractionation is closer to 1 than that
of other particle types (Table~\ref{tab:K_d}).
This result is consistent with \citet{geibert2004} whose laboratory study shows
no definitive affinity of either \chem{Th} or \chem{Pa} to biogenic silica.
Their average $K$ values actually suggest that thorium has a higher affinity to
biogenic silica.

We included lithogenic particles in the model from deposited dust.
Since lithogenic particles have a strong relative affinity with \chem{^{230}Th}
($K_{\chem{Pa},\mathrm{Lith}} / K_{\chem{Th},\mathrm{Lith}} = 0.2$) and since
they are mostly prevalent in the (northern) Atlantic Ocean, much of the
\chem{^{230}Th} is
scavenged in the Atlantic Ocean.
Calcium carbonate $K_\chem{Pa}/K_\chem{Th}$ is even smaller (0.024) and is
present througout much of the Atlantic but not in the Southern Ocean.
This leaves protactinium to be scavenged by biogenic
silica in the Southern Ocean when it arrives there through the AMOC\@.
Therefore, even with the smaller-than-conventional $K$ ratio of 0.4 for biogenic
silica, the familiar pattern of higher values of Pa/Th in the Southern Ocean
compared to most of the rest of the ocean is still reproduced.

\section{Conclusion}

The purpose of this study is two-fold.
Firstly, we wanted to set out a model of \chem{^{231}Pa} and \chem{^{230}Th}, complete with all the
necessary particle improvements and additions.
This includes an improved underlying particle dynamics from
NEMO--\textsc{Pisces} as well as the addition of lithogenic particles.
We have succeeded in this, and the model has been presented and implemented
such that it can be used for future studies, including for the modelling
and validation of ocean circulation in the past.

Secondly, this model helps us to study the interplay between particle and water
transport that control the \chem{^{231}Pa} and \chem{^{230}Th} profiles.
We have shown that by improving on the particles, we improved the
radionuclides, and we have shown to what extent the different particle phases
drive the scavenging of \chem{^{231}Pa} and \chem{^{230}Th}.

Dissolved \chem{^{230}Th} and \chem{^{231}Pa} concentrations are realistic in
intermediate-depth ocean (big improvement compared to \citet{dutay2009}) but not
in the surface: a finite equilibration time between the different radionuclide
phases may help.
The model can be extended to include this in future versions.
It would double the number of adsorption/desorption parameters whose values
should be determined through a careful literature and model sensitivity study.

In the deep ocean, the overestimation of \chem{^{230}Th} and \chem{^{231}Pa} in
several regions is likely to be caused, at least partly, by missing nepheloid
particles, but also manganese oxides (from hydrothermal vents) may improve the
distributions.
Additionally, the circulation may be too weak, not freshening AABW fast enough,
and hence \chem{[^{230}Th]}$_D$ and \chem{[^{231}Pa]}$_D$ become too high near the ocean floor.

\section{Code availability}

We have provided the model equations, source code and forcing files such that
our results can be reproduced.
Moreover, we encourage the reader to use our model, NEMO--ProThorP, to build extended and
improved models.
We have implemented two particle size classes of adsorbed nuclides, which is the
minimum that is needed to produce good results.
Instead of introducing a new tracer for every particle type, we only distinguish
between ``adsorbed onto small particles'' and ``adsorbed onto big particles'',
meaning that one is restricted, in this set-up, to include only particles that
sink with the two respective settling speeds for scavenging.
Whereas this is somewhat restricting, it is computationally efficient, and it
restricts the number of degrees of freedom (which is usually a good thing in
complex models).
NEMO--ProThorP can also be used to set up a non-instant equilibration model (with a $k_+$
and $k_-$) that may yield better surface values.

The underlying model, NEMO\;3.6 (svn revision 5283), can be downloaded from
\url{http://www.nemo-ocean.eu/} \citep{madec2016} after creating a login.
NEMO includes the biogeochemical \textsc{Pisces} model \citep{aumont2015}.
For reproducibility purposes, version 0.1.0 of the radionuclide and lithogenic particles model
code can be obtained at \url{https://dx.doi.org/10.5281/zenodo.1009065}
\citep{software::vanhulten2017:prothorp}.
However, we plan to make this code available as part of the NEMO model.

The NEMO model is available under the \href{http://www.cecill.info/}{CeCILL}
free software licence, modelled after the GNU~GPL\@.
The lithogenic particles and \chem{^{230}Th}/\chem{^{231}Pa}-specific code is licensed under the same
terms, or, at your option, under the
\href{https://www.gnu.org/copyleft/gpl.html}{GNU General Public License} version
3 or higher.
The authors do appreciate if, besides the legal adherence to copyleft and
attribution, they are informed about the use of the code.


\section{Author contributions}
The model and the simulations were designed by MvH, JCD and MRB\@.
MvH implemented the code and performed the simulations.
The manuscript was prepared by MvH with close collaboration and major
contributions from MRB and JCD\@.

The authors declare that they have no conflict of interest.

\begin{acknowledgements}
We would like to thank Marion Gehlen and Olivier Aumont for the discussions on
the calcite parameterisation in \textsc{Pisces}.
The first author is grateful to J\"org Lippold for both the discussion and
encouragement that also helped this work.

This study was supported by a Swedish Research Council grant (349-2012-6287) in
the framework of the French--Swedish cooperation in the common research training
programme in the climate, environment and energy agreement between VR and LSCE,
for the project ``Particle transport derived from isotope tracers and its impact
on ocean biogeochemistry: a \textsc{Geotraces} project in the Arctic Ocean''.
This study was partly supported by the project ``Overturning circulation and its
implications for the global carbon cycle in coupled models'' (ORGANIC, The Research
Council of Norway, grant no.\ 239965).

The authors wish to acknowledge the use of
\href{http://www.ferret.noaa.gov/}{Ferret}, a product of
\href{http://www.noaa.gov/}{NOAA}'s Pacific Marine Environmental Laboratory.
The plots in this paper were created by the Ferret visualisation library
ComPlot (\url{http://www.nongnu.org/complot/}) \citep{vanhulten2017:complot}.
\end{acknowledgements}

%
\DeclareRobustCommand{\DutchName}[4]{#1,~#3~#4}

\bibliography{klimato_artikoloj,klimato_datumoj,klimato_libroj,klimato_prelegoj,klimato_software,klimato_tezoj}{}
\bibliographystyle{apalike}
\end{document}